\documentclass[pdflatex,sn-mathphys-num]{sn-jnl}% Math and Physical Sciences Numbered Reference Style
\usepackage{graphicx}%
\usepackage{multirow}%
\usepackage{amsmath,amssymb,amsfonts}%
\usepackage{amsthm}%
\usepackage{mathrsfs}%
\usepackage[title]{appendix}%
\usepackage{xcolor}%
\usepackage{textcomp}%
\usepackage{manyfoot}%
\usepackage{booktabs}%
\usepackage{algorithm}%
\usepackage{algorithmicx}%
\usepackage{algpseudocode}%

\usepackage{listings}%
\theoremstyle{thmstyleone}%
\theoremstyle{thmstyletwo}%

\theoremstyle{thmstylethree}%

\begin{document}

\title[Article Title]{Decentralized Stratified Sampling for Low-Latency Approximate Geospatial Data Stream Processing in Edge-Cloud Architectures}


\author*[1]{\fnm{Isam Mashhour} \sur{Al Jawarneh}}\email{ijawarneh@sharjah.ac.ae}

\author[2]{\fnm{Lorenzo} \sur{Felletti}}\email{lorenzo.felletti@studio.unibo.it}

\author[2]{\fnm{Luca} \sur{Foschini}}\email{luca.foschini@unibo.it}

\author*[2]{\fnm{Paolo} \sur{Bellavista}}\email{paolo.bellavista@unibo.it}

\affil[1]{\orgdiv{Department of Computer Science}, \orgname{University of Sharjah}, \orgaddress{\city{Sharjah}, \postcode{27272}, \country{United Arab Emirates}}}

\affil[2]{\orgdiv{Dipartimento di Informatica – Scienza e Ingegneria}, \orgname{University of Bologna}, \orgaddress{\street{Viale Risorgimento 2}, \postcode{40136}, \city{Bologna},  \country{Italy}}}

\abstract{The exponential growth of geospatial data streams flowing from IoT devices challenges conventional cloud-based analytics, which typically suffer from network bandwidth waste and latency, basically attributed to the data being managed completely by Cloud, such as centralized sampling. To address this gap, we propose EdgeApproxGeo, a novel edge-cloud architecture that performs spatial-stratified online sampling at network edge devices near data sources. Our system introduces a novel sampling method called EdgeSOS, which is a unique decentralized, geohash-based stratified sampling algorithm designed to operate independently at resource-constrained edge nodes without cross-node synchronization, coupled with spatial-aware data distribution and topic routing in Apache Kafka data stream ingestion, aiming at optimizing downstream data stream processing analytics. We evaluated our system on two real-world geo-referenced datasets, mobility and air quality, and EdgeApproxGeo achieves a significant speedup over cloud-only baselines while maintaining errors in check (e.g., MAPE $<$ 10\% error rate at 80\% sampling rate). We further demonstrate that coarser geohash granularity (e.g., Geohash-5) can reduce error figures by 30\% as compared to finer counterparts (i.e., Geohash-6), thus revealing a tunable accuracy-efficiency trade-off. Our standard-compliant prototype, built atop Apache Kafka and Apache Spark, further validates the utility of edge-deployed approximate query processing for real-time big geospatial data analytics.}



\maketitle
\section{Introduction}\label{sec1}

The ubiquitous availability of GPS-enabled IoT devices, from vehicle telematics to environmental sensors, is feeding petabyte-scale daily georeferenced data streams  \cite{ConnectedVehicleData24, GHAFFARPASAND2024103815, FogCloudInterpolation25,UTrack25}. Location-based services (LBSs), such as real-time traffic optimization, environmental monitoring, and urban planning, are highly dependent on timely big data analytics that utilize those streams \cite{al2020locality, al2020efficient, HUANG2024}. Nevertheless, conventional exact query processing on full datasets is computationally expensive and performs poorly, especially during spikes in data traffic, thus deteriorating latency Service Level Objectives (SLOs). Approximate Query Processing (AQP) therefore offers a natural solution that trades marginal accuracy for significant efficiency gains \cite{wei2019online, li2018approximate}.

Recent research efforts have focused on exploring AQP in cloud-based and edge-based approaches; however, a clear research gap remains for geospatial big data stream analytics. On the one hand, cloud-based systems such as  SAOS \cite{al2019spatial}, ex-SAOS \cite{al2020spatially}, and SpatialSSJP \cite{al2023spatialssjp} have shown the demonstrated  utility of \textit{stratified sampling} using Geohash-based spatial splitting to preserve geospatial statistical representativeness. However, those systems are fundamentally cloud-based, which means that all raw data is migrated to a central computing cluster where sampling is performed. This \textit{transfer-then-filter} approach consumes costly network bandwidth for transmitting data which will eventually be discarded, therefore introducing unnecessary latency, even before beginning any computation and that is considered as a critical drawback in the current arena of IoT deployments.

Edge-based Stream Processing Engines (SPEs), such as EdgeWise \cite{fu2019edgewise} and AgileDART \cite{ching2025agiledart} focus on resource competition and scalability in resource-constrained edge devices to process fast-arriving data streams. EdgeWise works by replacing plain OS schedulers with a congestion-aware engine-level scheduler to optimize task execution order. AgileDART employs a decentralized data processing architecture and a bandit-based path planning approach to optimize data shuffling. Those systems primarily aim to balance QoS goals by maximizing throughput and minimizing end-to-end latency. They are considered efficient for general-purpose big data stream processing; however, they are unaware of the domain-specific structure of geospatial big data. Consequently, they cannot perform geospatial-aware operations, such as geo stratified sampling or spatially-aware upstream to downstream data routing and forwarding.

In the same vein, other edge computing systems in the literature focus more on data movement challenges. For instance, AggNet \cite{kumar2021aggnet} proposes a multi-tier aggregation network that spans edge, transit, and cloud data centers, aiming at minimizing the underlying cost of WAN data traffic, achieving this by strategically placing aggregation points. However, AggNet’s aggregation depends on applying standard deterministic aggregation operators (e.g., sum, count) that do not encompass statistically plausible and domain-aware sampling approaches that is required for geospatial data. In other words, they mainly work on optimizing data reduction by applying deterministic aggregation; however, they typically do not apply probabilistic sampling strategies that focus on preserving the underlying data distribution that is typically required for error-bounded geospatial big data analytics.

In the recent literature, relevant work includes those efforts that focus on approximate computing at edge devices. For example, ApproxECIoT \cite{zhang2022novel} aims to design an adaptive sampling strategy by proactively changing stratum sizes based on the observed statistical variance in characteristic values. However, its main focus is the localized sampling design and error control at the edge. The sampling method operates independently on each distributed edge node without employing a globally coordinated spatial stratification mechanism. As a consequence, it does not support global spatial representativeness between the participating distributed edge computing nodes, which mostly causes biased statistical computations (i.e., aggregates) when answering geospatial queries, which otherwise require a uniform coverage of the entire geographical region.

Most importantly, domain-specific systems such as GeoEkuiper \cite{huang2024geoekuiper} emphasize the requirement to consider geospatial characteristics (e.g., OGC-compliant SQL functions) in edge SPEs. Introduces a cloud-cooperated mechanism to dynamically transfer complex computational tasks between the edge and the cloud nodes, aiming at optimizing the throughput. However, data reduction is not the main focus in this work; it does not incorporate a new edge-hosted sampling method that is designed specifically to preserve the statistical utility of geospatial data streams.

The picture that emerges from the recent literature is the following, cloud-based SPEs consider geospatial statistical characteristics and incorporate functionalities for geospatial statistical modeling, they however,  neglect the network cost associated with transferring raw geospatial data streams to the Cloud layer. On the other hand, edge-based SPEs are optimized for network efficiency, utilization, and compute resources; however, they are unaware of geospatial data characteristics. Thus, there is a research gap that calls for designing a system that combines both functionalities (i.e., Cloud and edge ) without their limitations, namely, a system that performs statistically plausible, geospatial-aware data sampling at the network resource-constrained edge devices, and thereafter serves the reduced representative data to a cloud deployment, aiming at optimizing the entire end-to-end data processing pipeline.

To close this gap, this paper proposes EdgeApproxGeo, as a novel edge-cloud architecture for approximate geospatial analytics. Our work makes two main contributions:
\begin{enumerate}
    \item  We have designed EdgeSOS (Edge-based Spatial-aware Online Sampling) as a unique decentralized, geohash-based stratified sampling algorithm specifically for resource-constrained edge nodes. EdgeSOS works in a distributed and synchronization-free manner, which guarantees that each edge computing node independently produces a spatially representative sample.
    \item We introduce a new Kafka-based geospatial-aware data distribution mechanism which partitions spatial data streams by neighborhood (which is derived from a geohash-to-polygon mapping). Our design guarantees that the edge-sampled, spatially-partitioned data streams are delivered directly to appropriate Apache Spark worker nodes and partitions, thus significantly minimizing costly network shuffling and thereby enabling efficient parallel aggregation.
\end{enumerate}

To demonstrate the utility of our novel system, we have built a  prototype of our EdgeApproxGeo system on top of Apache Kafka and Spark (the two main systems for ingestion and processing of large data streams). We have performed extensive experiments using a real-world large geospatially-tagged mobility and hyperlocal Air Quality (AQ), and the results show that EdgeApproxGeo achieves a 1.2$\times$ speedup, on average, compared to cloud-only baselines (e.g. SpatialSSJP \cite{al2023spatialssjp}), while maintaining statistically bounded error (e.g. MAPE circa 7–10\% at an 80\% sampling fraction depending on the geohash granularity). Our results also show that our system supports a tunable accuracy-efficiency trade-off; for example, coarser Geohash granularities (e.g., Geohash-5) reduce error by approximately 30\% to 33\% compared to finer counterparts (e.g., Geohash-6), e.g., from circa 10.3\% to around 6.9\% MAPE at 80\% sampling fraction.

In summary, our novel system EdgeApproxGeo incorporates the benefits of cloud-based big data AQP with the resource-constrained designs of edge-based SPEs, without their limitations, aiming at enabling scalable, real-time big geospatial data analytics with QoS guarantees.

\section{Related Work}\label{sec2}

Challenges associated with processing fast-arriving big geospatial data streams have fueled significant research at the confluence of Approximate Query Processing (AQP), Edge Computing (EC), and distributed Data Stream Processing (DSP). Aiming at providing a concise, clear structured context of our contribution in this paper, we classify the related state-of-art based on two research directions: (1) the location where the approximation is performed, where we distinguish between Cloud-based and Edge-based systems, and (2) the main functional focus of the system, where we distinguish between general-purpose DSP and Geospatial-Aware DSP systems. This categorizes the literature into four distinct groups, which we discuss in detail hereafter. A concise comparison-based summary of these systems and their main utilities versus our work is summarized in Table~\ref{tab:related_work_comparison}.

(1) Cloud-based Geospatial AQP DSP systems. This vein of literature includes works that focus on Cloud-based AQP DSP, including sampling. Examples include our own systems SAOS \cite{al2019spatial}, ex-SAOS \cite{al2020spatially}, SpatialSSJP \cite{al2023spatialssjp}, and ApproxSSPS \cite{al2021qos}, which focus on Cloud-based stratified sampling for QoS-aware geospatial DSP. They typically operate by partitioning geospatial data streams into regularly-sized strata by utilizing Geohash-based geospatial data discretization. Those systems significantly reduce statistical estimation variance compared to Simple Random Sampling (SRS) \cite{thompson2012sampling}. However, since the sampling is performed on the cloud side, full data streams need to be shipped to the Cloud before sampling and processing. This can carry significant latency downstream as it requires excessive bandwidth, which means introducing additional latency. The work we present in this paper directly addresses this drawback, mainly by shifting the stratified sampling operations to Edge Computing (EC) devices near data sources.

(2) General-purpose edge SPEs. To mitigate the network latency bottleneck, a new group of SPEs has emerged to operate on resource-constrained edge computing nodes. State-of-the-art systems address the main challenges of resource competition and network dynamics based on several approaches. For example, EdgeWise \cite{fu2019edgewise} employs an approach that replaces plain OS schedulers with congestion-aware engine-level schedulers to optimize the task execution order on individual nodes, while AgileDART \cite{ching2025agiledart} employs a Distributed Hash 
Table-based  (DHT)-based peer-to-peer (P2P) overlay network architecture and a bandit-based exploitation-exploration path planning model to improve scalability and adapt to heterogeneous edge network dynamics. In the same vein, distributed big data stream processing frameworks such as NebulaStream \cite{michalke2025nebulastream, markl2024nebulastream} offer a robust platform for heterogeneous edge-cloud continuum. They utilize hardware-tailored query compilation and multi-query optimization to efficiently compose and execute complex data stream query graphs on volatile and resource-constrained edge-cloud infrastructures. However, those systems, which are adept at general-purpose DSP, are unaware of the domain-specific nature of data being processed. In more detail, they are unaware of the geospatial characteristics of data, and therefore are not designed to perform geo-aware operations like spatially-stratified sampling or data splitting and query routing. Our work complements those powerful systems by offering a domain-specific, statistically-plausible geospatial preprocessing layer, which can reduce the data stream served to those DSP engines.

(3) General-purpose edge-native big data AQP Systems.
The need to perform AQP on fast-arriving data streams at edge near the IoT has motivated the emergence of several general-purpose systems that are not intrinsically designed for geospatial data streams. For example, ApproxECIoT \cite{zhang2022novel} was designed to perform adaptive stratified reservoir sampling and perform delegated sub-computing tasks (e.g., statistical aggregations and intermediate results calculations) locally on data streams that pass through the edge nodes. It then serves the intermediate results to a cloud layer for further processing (e.g., final aggregation). Another example within the same consortium is Fossel \cite{abdullah2020fossel}, which applies reservoir sampling on a selected subset of fog nodes, then it runs queries on those nodes to reduce network communication and processing latency. Within the same category, works such as a data-aware edge sampling method introduced by \cite{wolfrath2020poster}, Adaptive Window-Based Sampling (AWBS) \cite{hafeez2020adaptive}, and efficient transmission of dependent data streams \cite{wolfrath2022efficient}, explore edge-based data sampling for aggregation queries; however, they do not consider spatial characteristics for data stratification or Geospatial AQP. While such systems show the robustness of edge-based AQP, they perform stratification based on non-spatial strata designs (e.g., by sensor ID or time window), thus rendering them unable to preserve the spatial distribution characteristics of the raw data. Our work aims at filling this gap by applying a geospatial-aware geohash-based stratification at EC nodes.

(4) Geospatial-aware Edge-native DSP systems.
Those are the systems that consider the geospatial characteristics of the data streams arriving at the edge nodes. An example from the literature is a system called GeoEkuiper \cite{huang2024geoekuiper}, which is a domain-specific system designed to host geospatial operations (e.g., open geospatial consortium (OGC)-compliant SQL functions) as an integral part of an end-to-end edge-native SPE. It employs a rule-oriented, cloud-assisted framework to run complex geospatial queries. However, its main focus is not on reducing the size of the data stream at the edge, and it does not feature a specialized edge-native geospatial sampling method to preserve the statistical properties of geospatial data streams. Other works within the same category include AggNet \cite{kumar2021aggnet}, which is a cost-effective data aggregation network that operates cross-over in end-to-end edge and cloud pipelines, with the aim of minimizing WAN data traffic by reducing WAN network traffic between IoT devices and the cloud layer. However, its aggregation strategy is limited to simple statistics-based data reduction methods (e.g. sum, count) and does not encompass advanced sampling techniques that preserve the analytical value of the data as they are in the IoT source. Moreover, systems such as StreamSight \cite{georgiou2018streamsight} employ a weighted hierarchical sampling approach; however, those are tightly coupled with specific query models and do not have out-of-the-box support for geospatial DSP.

Other edge and cloud-based approaches in the recent literature include a Sketching, such as a work by \cite{wu2020sketching}, which employs distributed sketching for real-time statistics computation in the cloud environments. In addition, \cite{alencar2020fot} introduces FoT-Stream that takes advantage of wavelet transforms and concept drift to perform fog-based data reduction. In the same vein, other systems such as EdgeStreaming \cite{ye2025edgestreaming} and RealEdgeStream (RES) \cite{ali2020res} focus on task offloading and video analytics, respectively. However, these works do not incorporate geospatial stratification awareness or address the specific statistical requirements of geospatial AQP.

Current works of recent state-of-the-art are either cloud-based systems (Category 1), which incorporate modules for comprehending geospatial statistical characteristics, but do not consider the cost on the network  associated with transferring raw data streams , or edge-based systems (Categories 2, 3, and 4), which are engineered for optimized network and compute resources optimization;  however, they are either unaware of geospatial characteristics of the data stream, employ non-spatial stratification, or lack dedicated, statistically-plausible sampling approaches engineered specifically for EC. Furthermore, most geospatial-aware EC systems are unable to incorporate spatially-aware data sampling at the edge with efficient downstream data processing in the cloud. This constitutes an obvious gap in the literature, where there is a need for a system that performs statistically plausible, geospatially-aware data sampling at the edge network, and thereafter forwards the sampled data to cloud downstream, aiming ultimately at optimizing the entire end-to-end SPE pipeline. Our work in this paper fills this critical gap by introducing EdgeSOS, which is, to the best of our knowledge, the first edge-native, geospatial-stratified online sampling method, in addition to incorporating a new spatial-aware topic routing module to guarantee efficient downstream big data analytics, these constitute two pillar components that are intrinsically incorporated in our EdgeApproxGeo system and work synergistically to achieve its design goals.

\begin{sidewaystable*}[t]
\centering
\caption{Detailed comparison of related work with EdgeApproxGeo.}
\label{tab:related_work_comparison}
\footnotesize 
\begin{tabular}{@{}lcccccc@{}}
\toprule
\textbf{System / Work} & \textbf{Deployment} & \textbf{SS}\footnotemark[1] & \textbf{DS}\footnotemark[2] & \textbf{AQP} & \textbf{Primary Focus} & \textbf{Key Limitation} \\
\midrule
SAOS / SpatialSSJP \cite{al2019spatial, al2023spatialssjp} & Cloud & Yes & No & No & Geospatial AQP & "transfer-then-filter" model \\
AgileDART \cite{ching2025agiledart} & Edge & No & No & No & General SPE & does not support spatial/AQP \\
NebulaStream \cite{michalke2025nebulastream} & Edge/Cloud & No & No & No & Distributed SPE & No built-in Spatial/AQP \\
ApproxECIoT \cite{zhang2022novel} & Edge/Cloud & No & Yes & Yes & General AQP & Non-spatial strata \\
Fossel \cite{abdullah2020fossel} & Fog & No & No & Yes & Latency Reduction & Centralized node selection; Generic sampling \\ 
AWBS \cite{hafeez2020adaptive} & Edge & No & Yes & No & Adaptive Windows & No spatial awareness\\
Data-Aware Edge Sampling \cite{wolfrath2020poster} & Edge & No & Yes & No & Aggregate Queries & No spatial stratification \\
AggNet \cite{kumar2021aggnet} & Edge/Cloud & No & No & No & Cost-Aware Aggregation & No AQP support; Exact aggregation only \\ 
GeoEkuiper \cite{huang2024geoekuiper} & Edge/Cloud & No & No & No & Geospatial SPE & No sampling strategy; Exact processing only \\
StreamSight \cite{georgiou2018streamsight} & Edge/Cloud & No & No & Yes & Query-Driven AQP & Centrally planned sampling; No native spatial support \\ 
Sketching \cite{wu2020sketching} & Cloud & No & Yes & Yes & Real-time Statistics & Not for spatial data \\ 
FoT-Stream \cite{alencar2020fot} & Fog & No & No & Yes & General IoT Streaming & Time-based approximation; No spatial AQP \\
EdgeStreaming \cite{ye2025edgestreaming} & Edge & No & No & No & Secure Streaming & Not geospatial-focused; No sampling \\
RES \cite{ali2020res} & Edge/Cloud & No & No & No & Video Analytics & Not geospatial-focused; Exact processing \\
\textbf{EdgeApproxGeo (Ours)} & \textbf{Edge/Cloud} & \textbf{Yes} & \textbf{Yes} & \textbf{Yes} & \textbf{Geospatial AQP} & \textbf{N/A} \\
\bottomrule
\end{tabular}
% Footnotes defining the abbreviations
\footnotetext[1]{SS: Spatial Strata (specifically for sampling).}
\footnotetext[2]{DS: Decentralized Sampling (autonomous edge decision-making).}
\end{sidewaystable*}

\section{EdgeApproxGeo: Edge-assisted architecture for the efficient processing of big geo-referenced data streams}\label{sec3}

\begin{figure}[h]
    \centering
    \includegraphics[width=0.90\linewidth]{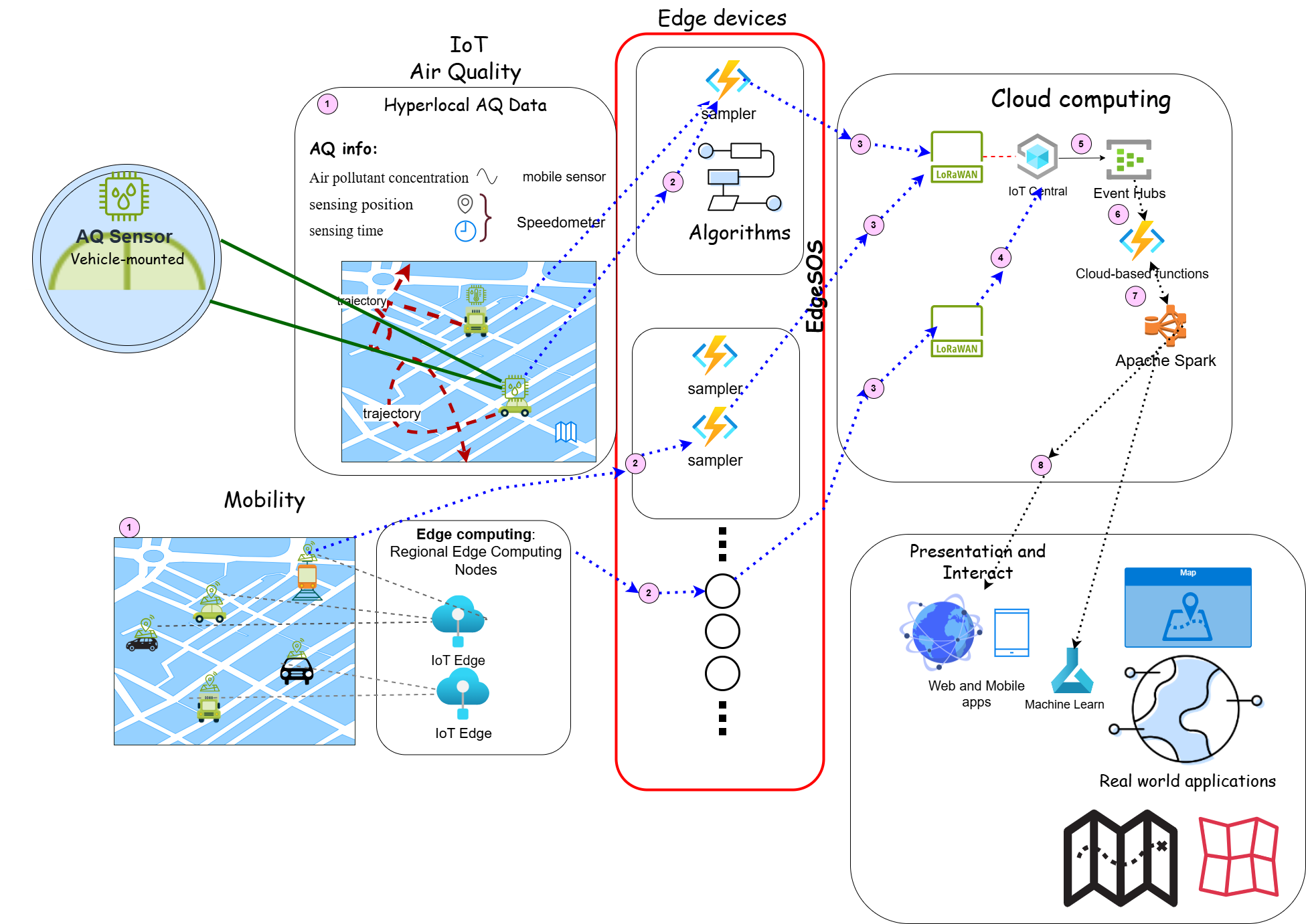}
    \caption{A high-level architecture of EdgeApproxGeo. IoT devices stream data to edge computing nodes, which then apply EdgeSOS sampling locally and a spatial-aware routing before forwarding the data to Kafka. Sampled data is then served to Apache Spark which performs approximate data analytics, and returns results with rigorous error bounds.}
    \label{fig:arch}
\end{figure}

\subsection{Problem statement and system model}\label{subsec:problem_statement}

The main goal of our system is to enable real-time, approximate big data analytics over huge, fast-arriving geospatial data streams that are generated by mobile IoT devices (e.g., vehicle telematics or low-cost air quality sensors mounted on moving vehicles and collecting hyperlocal AQ data). The main challenge that is typically encountered in such scenarios is to minimize the computational and heavy network loads typically imposed on the central cloud cluster deployments with minimal loss in statistical representativeness of the produced analytical results. We aim at supporting mainstream geo-statistical queries on big geospatial data streams, such as computing the average speed or count of vehicles per geohash over a tumbling time window.

Our system is engineered to operate under the following system model and assumptions, which are consistent with our evaluation setup (discussed shortly) and the real-world mobility and AQ big data:
\begin{itemize}
    \item \textbf{Data model}. We define the input data stream as a complete, unbounded sequence of georeferenced tuples ingested by the system from all data sources. A typical data tuple has the following structure (sensor\_id, timestamp, latitude, longitude, measurement), where sensor\_id is a unique identifier for the source (e.g., a specific taxi, or AQ sensor identification number). A sub-stream is defined as a subset of this multi-source data stream, and it normally flows from a single or multiple geographically-nearby logical data sources (i.e., a single sensor\_id, or multiple sensors that are co-located geospatially). For example, the continuous sequence of messages from one (or multiple co-located) taxis or hyperlocal AQ sensors is considered one sub-stream in our terms.
    
    \item \textbf{Spatial model}. Our geospatial model is structured as follows: the area of interest is divided into a regular grid of fixed-sized, adjacent, non-overlapping cells by using Geohash encoding. Each incoming data stream tuple is assigned to a single cell based on its (latitude, longitude) coordinates. In this context, data distribution refers to the statistical properties (e.g., mean, variance) of the measurements (like speed) within a given geohash cell over a time window. Although data sources are moving, the stratification is performed on the locations of readings (e.g., AQ, speed), not the ID of the data source (e.g., sensor\_ID). A single sub-stream (from one taxi), therefore, contributes tuples to several geohash strata as the source (e.g., sensor, vehicle) moves, and thereafter each of those strata is sampled independently.
    
    \item \textbf{System architecture}. We assume a three-tier edge-cloud architecture as follows: (1) IoT Devices (i.e., data sources) which generate raw georeferenced data streams, (2) Edge nodes, which are the resource-constrained devices, and are typically co-located with the IoT data sources, and (3) a Cloud cluster computing framework (e.g., Apache Spark) which performs the ultimate, computationally-intensive big data stream analytics. The communication and synergy between those three tiers is managed by a distributed messaging system (e.g., Apache Kafka).
\end{itemize}

We have designed our system EdgeApproxGeo (discussed shortly in subsection \ref{subsec:system_design}) so that it operates within the premises of this problem statement and system model. To achieve this, we aimed at designing a system that performs spatial-aware data reduction at the edge near the IoT data sources, and thereafter sends the samples to Cloud computing clusters for AQP, aiming at achieving optimized running and accuracy results over the entire end-to-end pipeline.

\subsection{EdgeApproxGeo system design and main components}\label{subsec:system_design}

to solve the problem declared in the previous subsection, we propose EdgeApproxGeo, as a new first-in-class edge-cloud computing architecture that encompasses two main components (Figure~\ref{fig:arch}) as follows:

\begin{enumerate}
    \item EdgeSOS (Edge-based Spatial-aware Online Sampling). An integral part of our system is a decentralized geohash-based stratified sampling module that operates on edge nodes near IoT sources. It ensures that each geohash cell (stratum in our system model definition, see subsection \ref{subsec:problem_statement}) is well-represented in the sample, thus avoiding situations that cause overlooking sparse regions (i.e., regions with fewer IoT readings) .
    \item Spatial-aware data distribution component. It encompasses a spatially aware big data routing module that partitions the edge-sampled data by neighborhood (which is a coarser aggregation of geohash values). This module operates on the interface between edge nodes and the distributed messaging system (Apache Kafka in our case). This design ensures that the data flowing to the cloud is, by then, already partitioned on a spatial key, thus minimizing costly network shuffles during Spark's analytics (e.g., aggregation of edge-sampled data).
\end{enumerate}

The following are the core principles that guide the design of our EdgeApproxGeo system: (1) Resource efficiency. Sampling is performed at the edge near the IoT, thus minimizing upstream bandwidth and offloading a significant load from the cloud. (2) Transparency. Front-end application developers submit standard SQL-like queries, whereas the heavy-lifting of complexities that involve data sampling, routing, and error estimation are handled automatically by the underlying components of our system. That is, SQL-like spatial queries are compiled down into efficient query plans that call geospatial-aware tools of our system in an end-to-end pipeline, while the logistic handling is hidden from front-end developers. (3) Adaptability. A module in our system comprises a feedback loop mechanism that can dynamically adjust sampling fractions for each edge node based on local resource availability and SLOs defined in the user query.

As depicted in Figure~\ref{fig:arch}, EdgeApproxGeo is designed based on a three-tier architecture as follows: (1) Edge data ingestion tier. IoT devices read and stream georeferenced data tuples to nearby edge computing nodes. (2) Edge sampling and routing tier. Each edge computing node applies EdgeSOS sampling locally to sample data stream tuples proportionally from each geohash stratum. It thereafter ships selected samples to Kafka topics (the distributed messaging system we utilize in our system) that are partitioned by neighborhood. (3) Cloud data analytics tier. Spark worker nodes are subscribed to messaging system topics, and thereby consume data that arrives from spatially-partitioned topics, and compute geo-statistics such as aggregates (e.g., average PM$_{10}$ per geohash), and eventually return results with rigorous statistical error bounds. Next, we start by discussing EdgeSOS, an integral component of our system.

\subsection{The EdgeSOS Algorithm}\label{subsec:edgesos_algo}

EdgeSOS is a stratified sampling component that operates independently on each EC node. It is analogous to Cloud-based geospatial stratified sampling methods that appear in the recent literature, specifically SAOS \cite{al2019spatial}, ex-SAOS \cite{al2020spatially}; however, we have redesigned and repurposed EdgeSOS so that it operates efficiently for resource-constrained EC devices.

The Algorithm~\ref{alg:edgesos} lists the operational mechanism of EdgeSOS. Each EC node in the network receives data stream input tuples that are flowing from a local area it is covering, it then splits the data stream into geohash-based strata (line 2), and thereafter computes the sample size for each stratum (line 3), and finally applies Simple Random Sampling (SRS) within each stratum independently (line 6). The union of all stratum samples is returned as the new sample to be forwarded to Cloud computing downstream (line 9).

\begin{algorithm}[H]
\caption{Edge-based Spatial-aware Online Sampling (EdgeSOS)}\label{alg:edgesos}
\begin{algorithmic}[1]
\Require messages: input stream of tuples; samplingFraction: target sampling ratio (e.g., 0.8 for 80\%)
\Ensure Sample: set of sampled tuples

\State Sample $\Leftarrow \emptyset$
\State SUB $\Leftarrow$ UpdateSub(messages) \Comment{Partition stream into geohash-based strata}
\State sizesMap $\Leftarrow$ specifySampleSize(samplingFraction, SUB) \Comment{Compute sample size per stratum}

\For{each stratum $g_i$ in SUB}
    \State $n_i \Leftarrow$ sizesMap[$g_i$] \Comment{Target sample size for stratum $g_i$}
    \State sample$_i$ $\Leftarrow$ SRS\_Sample($g_i$, $n_i$) \Comment{Simple Random Sampling within stratum}
    \State Sample $\Leftarrow$ Sample $\cup$ sample$_i$
\EndFor

\State \Return Sample
\end{algorithmic}
\end{algorithm}

\begin{figure}[h]
    \centering
    \includegraphics[width=0.90\linewidth]{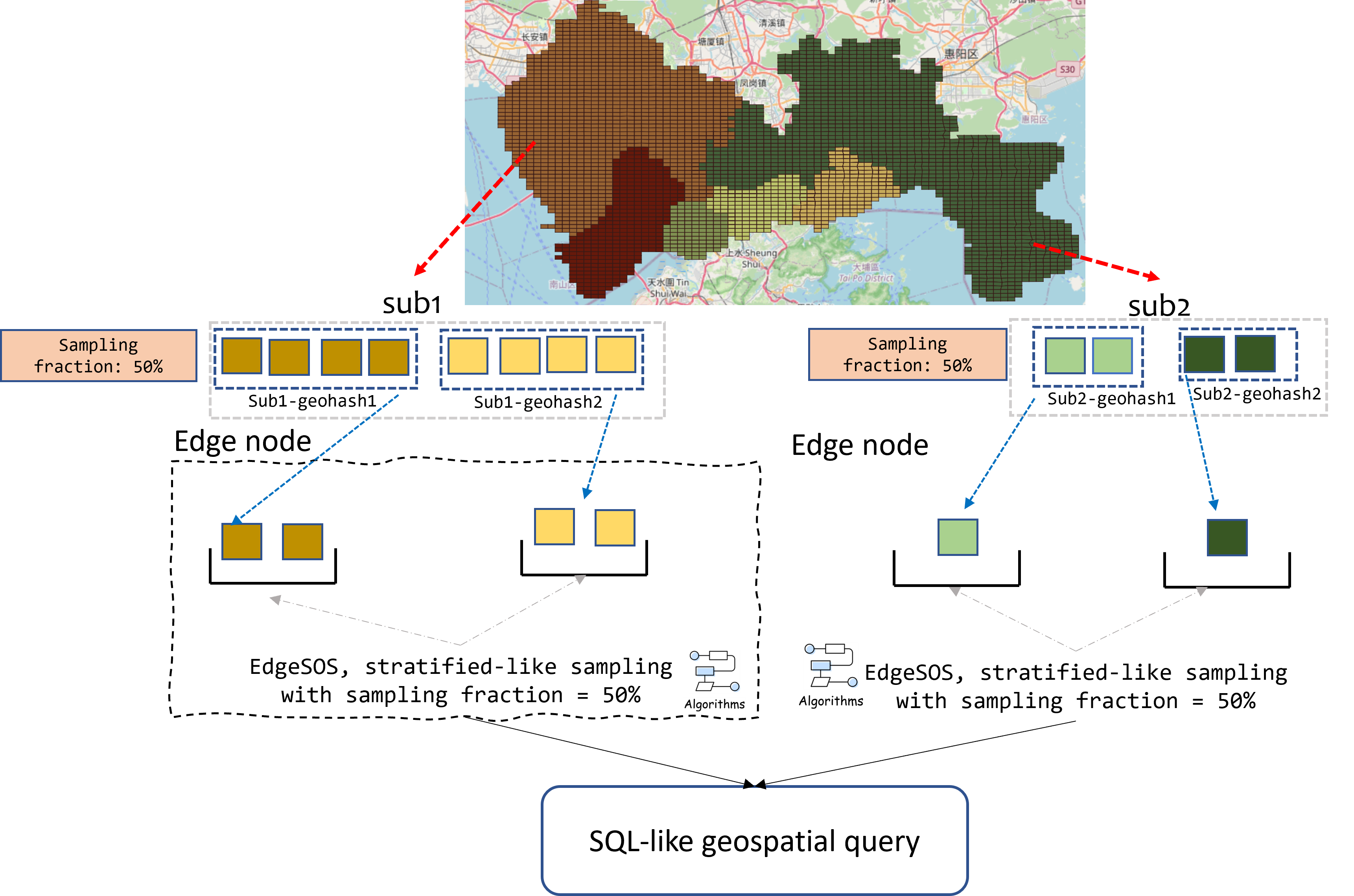}
    \caption{A toy example shows how an Edge computing node (that is covering a neighborhood) samples each geohash stratum independently. Algorithm~\ref{alg:edgesos} runs for each node at each time interval.}
    \label{fig:toy}
\end{figure}

Figure~\ref{fig:toy} illustrates how a single edge node samples items from each geohash stratum.

In the following subsection, we discuss some implementation optimizations that we obtain by using Rust systems-level programming language to implement EdgeSOS as part of our system.
\subsubsection{Implementation optimizations}

To guarantee to a plausible extent that EdgeSOS introduces only negligible latency to the end-to-end pipeline at the edge, we implemented it in the Rust systems programming language for several reasons including its performance as it compiles to native machine code, resulting in a performance that is comparable to C++ programming language. This choice has enabled us to introduce key optimizations in our design including the following: (1) Using Rust’s rayon crate, our sampler EdgeSOS performs sampling in a parallel fashion across geohash strata. This design has reduced latency significantly for data stream batches that contain $>$ 50K tuples (see subsection~\ref{EdgeSOSSampling} for more details). (2) We employ a precomputed inverted hashmap (geohash $\rightarrow$ neighborhood) for an efficient spatial mapping, which is employed specifically for neighborhood lookup (i.e., given a geohash, then the inverted hashing returns the enclosing neighborhood), thus enabling O(1) run time complexity for neighborhood lookups, and therefore avoiding computationally expensive point-in-polygon geo operations during runtime.
In the next subsection, we discuss the workflow which our system EdgeApproxGeo operates on.

\subsection{EdgeApproxGeo Workflow}\label{subsec4}

Algorithm~\ref{alg:approxgeo} lists the end-to-end workflow of our system EdgeApproxGeo. The system begins by accepting a geospatial continuous query (CQ) and a SLO resource budget (time and accuracy based). IoT data sources continuously generate big georeferenced data streams and push them to EC nodes. Every EC node receives data containing several geohashes covering its local area, then EdgeSOS is applied locally to sample data from each geohash independently, and sampled tuples are then forwarded to the cloud.

We assume a cost function that maps the geo CQ QoS constraints (i.e., accuracy and latency) into a sampling fraction that is applied to each edge node independently. Each node then performs a local sampling based on that fraction, and serves its data sample (representing a sub-stream) to a cloud-based compute node, once the Cloud-deployed component of the system receives all updates from all edge nodes, it runs the geospatial CQ query and returns a local result (e.g., average speed) with rigorous error bounds.

\begin{algorithm}[H]
\caption{EdgeApproxGeo overview}\label{alg:approxgeo}
\begin{algorithmic}[1]
\Require cquery: continuous query (Cloud-based node)
\Require runningBudg: overall budget (e.g., max latency 2s, max error 10\%)

\For{each time interval $t_i$}
    \State samplingFraction $\gets$ fractionCalc(runningBudg) \Comment{e.g., 0.8 for 80\%}
    \State stream $\gets$ receiveStream($t_i$)
    \While{!empty(stream)}
        \State tuples $\gets$ getTuples(stream)
        \State Sample $\gets$ EdgeSOS(tuples, samplingFraction) \Comment{Edge-based stratified sampling}
        \If{node is sampling} \Comment{Edge node}
            \State Send(parent, sample) \Comment{Forward to cloud}
        \Else \Comment{Cloud node}
            \State $\Theta \gets \Theta \cup$ sample
        \EndIf
        \State stream $\gets$ stream $\setminus$ tuples \Comment{Remove processed tuples}
    \EndWhile
    \State result $\gets$ run(cquery, $\Theta$) \Comment{Execute query in Spark}
    \State $e \gets$ computeError(result) \Comment{Compute error bounds (Section~\ref{subsec:error_estimation})}
    \State \textbf{output} result $\pm$ $e$
\EndFor
\end{algorithmic}
\end{algorithm}

In more detail, as apparent from the algorithm~\ref{alg:approxgeo} listing, the workflow of our system proceeds as follows: It first computes the sampling fraction based on the QoS constraints in the CQ (line 2), it then extracts data tuples for each sub-stream (few geohashes in this case) (line 4), thereafter it applies EdgeSOS on data received by each edge node independently (line 6), and it eventually forwards data samples to the cloud compute for further processing (line 8). The cloud node then aggregates data samples in $\Theta$ (line 10), and runs the geospatial CQ (line 14), in addition to computing the rigorous error bounds (line 15), and eventually returning both results (line 16). This process is repeated for each time interval using a tumbling-window paradigm of CQ processing.
In the upcoming subsection, we discuss some example CQs that are directly supported in our system in addition to others that can be composed based on those primitive queries.

\subsection{System geo-statistics (example queries)}\label{sec:geostats}

In our design, which resembles tree-based graph structures, the cloud-based node (i.e., the parent) receives data samples from all sub-streams as $\Theta$ and it then calculates geo-statistics (e.g., mean, sum). Since edge nodes sample data independently (i.e., without inter-node synchronization), the parent computes the estimated sum for one sub-stream $s$ (denoted as $\hat{t}_s$) using \eqref{eq:substream_sum} (i.e., a stratified sampling estimation within that node):

\begin{equation}
\hat{t}_s = \sum_{k=1}^{K_s} T_{s,k} = \sum_{k=1}^{K_s} N_{s,k} \cdot \bar{y}_{s,k}
\label{eq:substream_sum}
\end{equation}

where $\hat{t}_s$ is the estimated sum of the target variable in the sub-stream $s$, $T_{s,k}$ is a tuple in the sub-stream $s$ and the geohash stratum $k$,  $N_{s,k}$ is the population size of stratum $k$ within the sub-stream $s$, and $\bar{y}_{s,k}$ is the mean sample of the target variable in stratum $k$ within the sub-stream $s$. $K_s$ is the number of distinct geohash strata observed in the sub-stream $s$.

Assuming that we have $M$ sub-streams in the sampled data stream $\Theta$, then the estimated global sum of the target variable is calculated by aggregating the estimates from all sub-streams using \eqref{eq:total_sum}:

\begin{equation}
\widehat{\mathrm{SUM}}_\Theta = \sum_{s=1}^{M} \hat{t}_s
\label{eq:total_sum}
\end{equation}

Then, calculating the estimated population mean is straightforward. Let $I$ be the total number of unique geohash strata globally in all sub-streams. The estimated global mean of the target variable (denoted as $\overline{Y}_{\text{EdgeSOS}}$) is calculated by applying \eqref{eq:estimated_mean}:

\begin{equation}
\overline{Y}_{\text{EdgeSOS}} = \frac{\widehat{\mathrm{SUM}}_\Theta}{N_{total}} = \sum_{i=1}^{I} \left( \frac{N_i}{N_{total}} \right) \bar{y}_i
\label{eq:estimated_mean}
\end{equation}

where $\bar{y}_i$ is the sample mean of the target variable in stratum $i$,  $N_i$ is the population size of stratum $i$ (i.e., the total number of tuples in that geohash bracket). $N_{total} = \sum_{i=1}^{I} N_i$ is the total estimated population size (i.e., the total number of data tuples that arrived from the IoT stream at the time of calculation).

It is apparent that the global mean is the weighted average of the stratum means, where the weights are the proportion of the population in each stratum ($N_i / N_{total}$). $\bar{y}_i$ is an unbiased estimator of the true mean of stratum $i$ (attributed to employing Simple Random Sampling within each geohash independently), and $N_i$ is assumed to be known (or at least incrementally and accurately estimated, either using lightweight real-time counters or from similar data-history profiles). As a consequence, $\overline{Y}_{\text{EdgeSOS}}$ is thus considered as an unbiased estimator of the true global population mean.

\begin{figure}[h]
    \centering
    \includegraphics[width=0.90\linewidth]{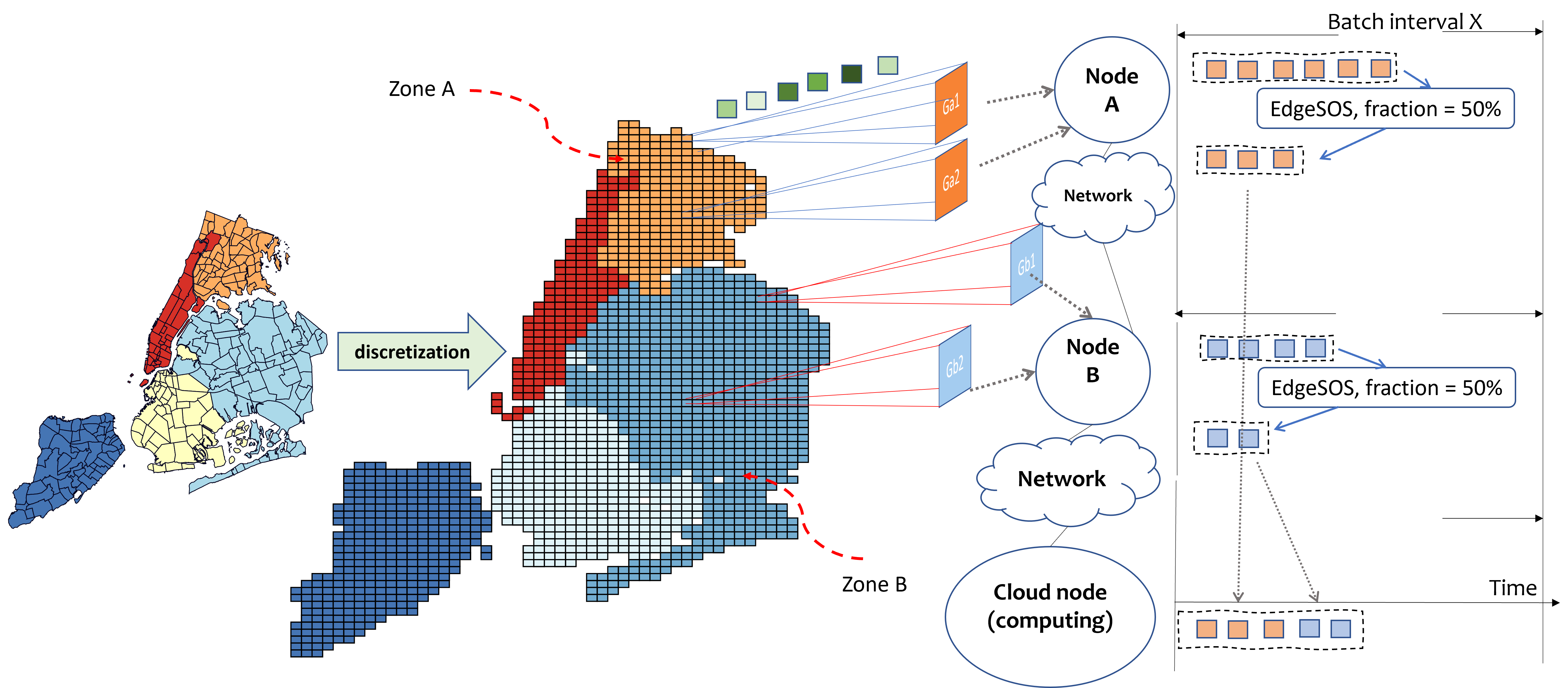}
    \caption{Edge nodes A and B perform decentralized sampling without inter-node coordination, each node samples its local strata (i.e., few geohash values) independently.}
    \label{fig:toy_nodes}
\end{figure}

Figure~\ref{fig:toy_nodes} shows a toy example, assume Edge nodes A and B sample data independently (without inter-node coordination). Imagine that 6 tuples from two geohash values $Ga1$ and $Ga2$ (two strata) arrived at node A (assume a sampling fraction of 50\%), then EdgeSOS sample 3 tuples in node A. Similarly, 2 out of 4 tuples are sampled from the second sub-stream arriving independently at edge node B (containing $Gb1$ (one stratum) and $Gb2$ (another stratum)). The sample size is 5 data tuples (3 from sub-stream data sampled at node A, and 2 from sub-stream data sampled at node B). Assume that attribute values of those sampled data tuples (e.g., average PM10 value) are (10,7,8) for data sampled at A, and (6,11) for those sampled at B, then the estimated sums are 25 and 17, respectively, and the Cloud-based grand total is 42, while the mean value is 8.4.
Since estimating geo-statistics is susceptible to errors, quantifying statistically the errors arising from the approximation is an essential part of our system, which is discussed in the next sub-section.

\subsection{Quantifying the uncertainty associated with sampling}\label{subsec:error_estimation}

In geo-statistics, estimating values of the target variable by relying on samples instead of population data intrinsically introduces some statistical uncertainty, which can be calculated. In our system EdgeApproxGeo, we introduce modules to quantify this kind of uncertainty relying on rigorous error bounds that are statistically derived from the stratified sampling theory \cite{thompson2012sampling}.

\subsubsection{Statistical foundations}

Let $ K $ be the total number of unique geohash strata that are observed globally at all edge nodes during a time interval. For each stratum $ k $ ($ 1 \leq k \leq K $): $ N_k $ represents the population size of stratum $ k $ (which is estimated via online counters or historical profiles). $ n_k $ is the number of tuples sampled from the stratum $ k $. $ y_{k,1}, \dots, y_{k,n_k} $ are the values of the target variable in the stratum sample $ k $.

The sample mean $\bar{y}_k$ and sample variance $s_k^2$ for each stratum $k$ are then calculated applying \eqref{eq:sample_mean_variance_stratum}:

\begin{equation}
\bar{y}_k = \frac{1}{n_k} \sum_{j=1}^{n_k} y_{k,j}, \quad s_k^2 = \frac{1}{n_k - 1} \sum_{j=1}^{n_k} (y_{k,j} - \bar{y}_k)^2
\label{eq:sample_mean_variance_stratum}
\end{equation}

Then using these stratum-level statistics, the approximate global sum of the overall $\widehat{\mathrm{SUM}}$ and the mean $\widehat{\mathrm{MEAN}}$ is calculated using the stratified sampling estimators defined in equation \eqref{eq:foundations_sum_mean}:

\begin{equation} 
\widehat{\mathrm{SUM}} = \sum_{k=1}^{K} N_k \cdot \bar{y}_k, \quad \widehat{\mathrm{MEAN}} = \frac{\widehat{\mathrm{SUM}}}{\sum_{k=1}^{K} N_k}
\label{eq:foundations_sum_mean}
\end{equation}

We next show how the system proceeds with error estimation by calculating the variance of these statistics.
\subsubsection{Variance estimation}

Because sampling is performed independently within each geohash stratum, the total variance of the estimated sum is calculated by computing the sum of the variances of all $K$ global strata (i.e., all geohash values globally) using \eqref{eq:var_sum}:

\begin{equation}
\widehat{\mathrm{Var}}(\widehat{\mathrm{SUM}}) = \sum_{k=1}^{K} N_k^2 \cdot \left(1 - \frac{n_k}{N_k}\right) \cdot \frac{s_k^2}{n_k}
\label{eq:var_sum}
\end{equation}

Subsequently, the variance of the estimated mean is calculated using \eqref{eq:var_mean}:

\begin{equation}
\widehat{\mathrm{Var}}(\widehat{\mathrm{MEAN}}) = \frac{1}{\left(\sum_{k=1}^{K} N_k\right)^2} \cdot \widehat{\mathrm{Var}}(\widehat{\mathrm{SUM}})
\label{eq:var_mean}
\end{equation}

Using these statistics, we next show a method to calculate the error bounds and confidence intervals, to quantify the uncertainty that is caused by the approximation (i.e., relying on samples instead of the population).

\subsubsection{Confidence intervals and error bounds}

Using the Central Limit Theorem \cite{thompson2012sampling}, our system constructs a confidence interval $ (1 - \alpha) \times 100\% $ applying \eqref{eq:confidence_interval}:

\begin{equation}
\widehat{\mathrm{MEAN}} \pm z_{\alpha/2} \cdot \sqrt{\widehat{\mathrm{Var}}(\widehat{\mathrm{MEAN}})}
\label{eq:confidence_interval}
\end{equation}

where $ z_{\alpha/2} $ is the upper $ \alpha/2 $ quantile of the standard normal distribution (e.g. $ z_{0.025} = 1.96 $ for 95\% CI).

In practice, our system reports the following.

\begin{equation}
\text{Approximate result} \pm \text{Margin of Error (MoE)}, where \quad \text{MoE} = z_{\alpha/2} \cdot \sqrt{\widehat{\mathrm{Var}}(\widehat{\mathrm{MEAN}})}
\end{equation}

In addition, a Relative Error (RE) for normalized interpretation is reported by calculating it using \eqref{eq:relative_error}:

\begin{equation}
\mathrm{RE} = \frac{\text{MoE}}{\widehat{\mathrm{MEAN}}} \times 100\%
\label{eq:relative_error}
\end{equation}

In the next subsection, we briefly discuss the approach by which our system reports this uncertainty rigorously to support AQP on fast arriving geo-referenced data streams.

\subsubsection{Practical implementation in EdgeApproxGeo and Cloud-side aggregation workflow}

Our system supports a dynamic operation approach and considers the nature of varying network conditions. In more detail, EdgeApproxGeo supports two modes of operation for data transmission from the edge to the cloud layer as follows: (1) Raw sampled data transmission mode. Edge nodes ship sampled raw tuples (that are augmented with Geohash values) to the Cloud layer. In this mode of operation, the Spark cluster performs all statistical aggregations (i.e., mean, variance, count) by grouping tuples by Geohash values and applying equations \eqref{eq:total_sum}, \eqref{eq:estimated_mean}, \eqref{eq:var_sum}, and \eqref{eq:relative_error} directly on the distributed dataset. (2) Pre-aggregated statistics transmission mode. To reduce network bandwidth, edge nodes can compute local stratum-level statistics (i.e., $\bar{y}_k$, $s_k^2$, $n_k$) for each geohash $k$. These statistics, together with population estimates $N_k$ (which is derived from lightweight online counters or historical profiles), are served to the Cloud layer for further processing. Spark then aggregates these partial statistics on all nodes to compute the global $\widehat{\mathrm{SUM}}$, $\widehat{\mathrm{MEAN}}$, in addition to variances using equations \eqref{eq:foundations_sum_mean}, \eqref{eq:var_sum}, and \eqref{eq:var_mean}.
For both cases, the final approximation results are then served with rigorous confidence intervals and error bounds (e.g., "Average speed = 32.4 km/h $\pm$ 1.8 km / h (95\% CI)") by applying equations \eqref{eq:confidence_interval} -- \eqref{eq:relative_error}. 

Finally, if the Relative Error (RE) exceeds a pre-specified threshold, a feedback loop triggers an adaptive sampling mechanism. This mechanism dynamically adapts the sampling fraction for subsequent micro-batch intervals to meet the QoS requirements that are specified in the continuous query's (CQ) Service Level Objectives (SLOs).

This design guarantees that the decentralized nature of data ingestion does not compromise the statistical rigor of the approximation. The geohash value is used for aggregation, and this allows EdgeApproxGeo to guaranty an unbiased estimation and a correct weighting of sparse versus dense regions, thus effectively applying the stratified sampling theory in a scalable and distributed manner.
\section{Implementation insights}\label{sec4}

\begin{figure}[h]
    \centering
    \includegraphics[width=0.90\linewidth]{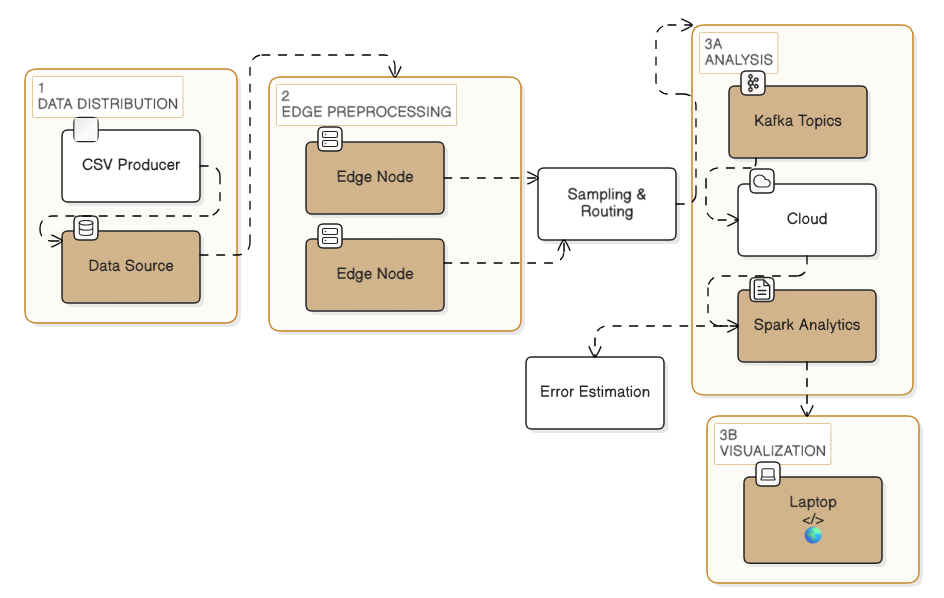}
    \caption{Component diagram of EdgeApproxGeo prototype. Data flows from CSV producer to Edge nodes (which perform sampling and routing) to Kafka topics that are then shipped to Cloud component, then Apache Spark on the Cloud (performs analytics on sampled data and serves error estimation).}
    \label{fig:flow}
\end{figure}

To show the utility of our system, we have designed and implemented a standard-compliant prototype of EdgeApproxGeo using two de facto standard platforms for Edge and Cloud computing, i.e., Apache Kafka and Apache Spark Structured Streaming. This technology stack was selected carefully because of their robust architectural properties, which are essential to realizing the main components of our system, including, most importantly, decentralized, geospatial-stratified sampling at the edge computing networks, in addition to spatial-aware data routing and partitioning. As depicted in Figure~\ref{fig:flow}, our system comprises fully implemented modules. In this section, we discuss the details of our implementation that is a containerized deployment, which utilizes the unique traits of Apache Kafka and Apache Spark to support our system's design. In the next sub-section, we briefly discuss the technology stack that we choose and the rationale in the context of application to implement our system main functionalities.

\subsection{ A purpose-built technology stack foundation}
\label{subsec:tech_stack}

Our choice of Apache Kafka and Apache Spark is driven by their ability to directly support the two core functionalities of our system EdgeApproxGeo, as  follows: (1) Apache Kafka~\cite{kreps2011kafka}, we selected it to implement our spatial-aware data distribution and routing component. The plain version of Kafka features a partitioned topic model that we adapt to implement the topic routing modules of our system. Specifically, we extend the plain Kafka topic partitioning and routing module that operates as follows: our patches create one Kafka topic for each neighborhood, and edge nodes publish their sampled data stream tuples directly to the matching topic (assuming that each neighborhood is served by one edge node). This guarantees that sub-stream data from each neighborhood is shipped to one matching Kafka topic. Kafka then continuously emits the topic arriving data stream tuples into a Cloud-based Spark deployment. This design guarantees, and ensures that Spark executors (each worker node in Spark hosts several executors) that consume from those topics receive data that are already geospatially pre-partitioned by geographic region (i.e., neighborhood) . This has a significant performance impact on subsequent stages of operations, as it reduces or even eliminates the need for computationally expensive network data shuffles during subsequent data aggregation stages, which constitutes a significant optimization that is unattainable with simple pub/sub brokers design, which typically lacks a partitioned log structure that is similar to that of Kafka. (2) For Cloud-based geospatial AQP on data stream tuples arriving from edge devices, we choose Apache Spark Structured Streaming ~\cite{armbrust2018structured, zaharia2010spark, zaharia2016apache} to implement our statistical integration and QoS loop feedback mechanism. Some of the overarching features of Spark Structured Streaming include its unbounded table abstraction, which treats data tuples flowing from edge nodes in the upstream containing pre-sampled, pre-partitioned data tuples as a continuous dataset. This design simplifies the process of computing per-geohash sample statistics (e.g. $\bar{y}_k$, $s_k^2$) as rolling aggregations within the premises of micro-batch data stream processing abstraction. Subsequently, Spark’s Catalyst optimizer seamlessly combines these data statistics that are obtained using our repurposed stratified-like sampling formulas to produce global confidence intervals, which can be achieved seamlessly using a single declarative SQL-like continuous query on the unbounded data stream. Additionally, Spark supports reporting these error metrics in real-time, which enabled us to introduce a lightweight patch to handle the QoS-aware loop-feedback module of our system, thus allowing our system to interactively adjust the edge sampling fractions to meet QoS targets defined as part of the SQL-like continuous query on the data stream.

This synergy of employing Apache Kafka for spatial orchestration (routing and pre-partitioning) and Apache Spark for the geo-statistical computations constitutes the backbone of the architectural design of our system EdgeApproxGeo, thus enabling an end-to-end workflow that is highly efficient and rigorous statistically. In the next subsection, we discuss how we deployed those two components to handle real case scenarios efficiently.

\subsection{EdgeApproxGeo implementation details}

We utilize Docker containers to deploy our prototype, thus enabling consistent, reproducible execution across edge and cloud computing environments. The deployment is composed of two main computing clusters as follows: (1) Kafka Cluster. We deploy two broker containers that are managed by a Zookeeper instance that is basically responsible for handling the creation of Kafka topics, the partitioning of arriving data tuples, and the routing of messages. (2) Spark Cluster. We deploy a master-worker topology (currently consisting of one master and one worker node), and is basically responsible for processing incoming data streams (pre-sampled tuples flowing continuously from edge nodes) and computing approximated aggregates (i.e., global sum or average).

\begin{figure}[h]
    \centering
    \includegraphics[width=0.90\linewidth]{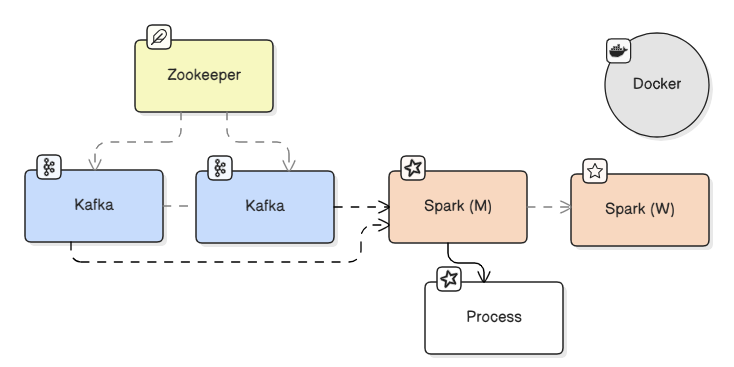}
    \caption{Container deployment structure. Five containers: 2 Kafka, 1 Zookeeper, 1 Spark Master, 1 Spark Worker. Dashed lines indicate communication via Kafka topics.}
    \label{fig:cluster}
\end{figure}

Figure~\ref{fig:cluster} shows the structure of the container deployment that we used to prototype our system. Communication between the components of the system occurs through Kafka topics exclusively, thus there is no direct inter-container networking. We opt for this decoupled design to enhance the scalability of our system.

The system encompasses three main components as follows: (1) Data distribution node. This node is employed as a proxy to simulate real-world IoT data streams real-time arrival, achieved by reading disk-resident data from a CSV file and then distributing the CSV file records to edge computing nodes through partitioned Kafka topics. Every edge computing node then consumes a specific Kafka topic partition, thus ensuring a balanced load. This design resembles geographic data sharding, where each partition corresponds to a spatial region or the coverage area of an edge node.

    \begin{figure}[h]
        \centering
        \includegraphics[width=0.90\linewidth]{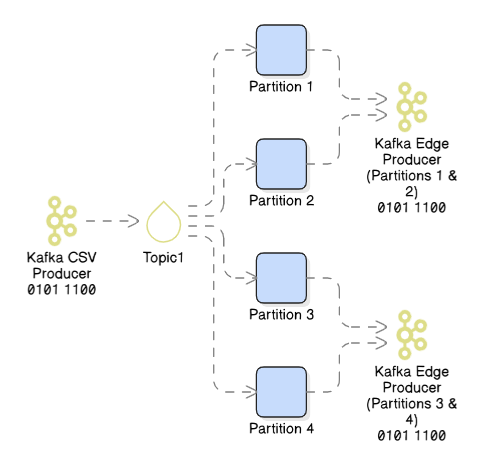}
        \caption{Data distribution via partitioned Kafka topic. Edge nodes consume from assigned partitions, thus ensuring load balance.}
        \label{fig:csv}
    \end{figure}

    (2) Edge data processing node. This constitutes the core functionality supported by our system, EdgeApproxGeo. Each edge computing node performs the following tasks: (A) consumes raw data tuples from its assigned Kafka partition. (B) It then computes a 6-character geohash for each data tuple it receives and then maps it to a corresponding neighborhood using a precomputed inverted hashmap (O(1) lookup functionality, which is supported by our design). (C) Therefore, our EdgeSOS sampling algorithm (Algorithm~\ref{alg:edgesos}) is applied to perform efficient decentralized stratified sampling for each geohash stratum individually. (D) Eventually, it routes sampled data tuples to neighborhood-partitioned Kafka output topics, thereby preserving spatial co-locality for downstream Spark processing in the Cloud.

\begin{figure}[h]
        \centering
        \includegraphics[width=0.90\linewidth]{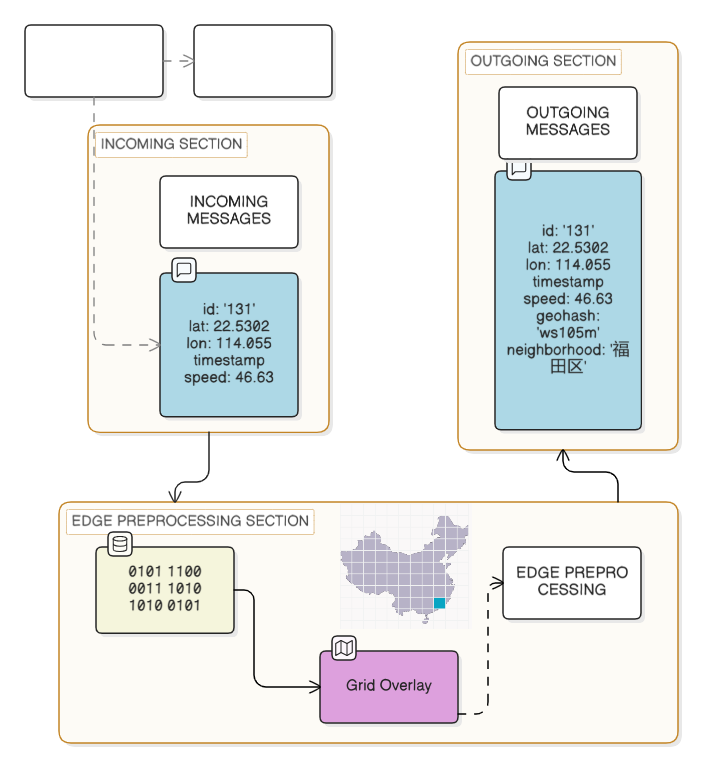}
        \caption{Message structure before (left) and after (right) edge processing. Added fields: \texttt{geohash}, \texttt{neighborhood}.}
        \label{fig:message}
    \end{figure}

   We implemented this important component in the Rust programming language for its performance benefits, where the edge binary leverages rayon for parallel sampling and FxHash for efficient geohash-keyed neighborhood lookups. This simplifies the enrichment of the message structure with geohash and neighborhood fields, as shown in Figure~\ref{fig:message}.

    (3) Cloud data analysis node. As an integral part of the main operation of our system, we introduced patches to the Cloud compute node hosting Apache Spark. It consumes data streams arriving from spatially-partitioned Kafka topics, it then computes approximate AQP statistical aggregates (e.g., mean speed for each geohash or neighborhood), and thereafter it estimates statistical error bounds by employing equations~\eqref{eq:var_sum}--\eqref{eq:relative_error}. We also incorporated an adaptive feedback loop mechanism to monitor relative error (RE), in such a way that if the RE exceeds a pre-defined threshold, it triggers a dynamic adjustment of the data stream sampling fraction at edge compute nodes.

\section{Results and discussion}\label{Results}
In this section, we present a detailed experimental evaluation of our system, EdgeApproxGeo, which represents our novel edge-cloud architecture for approximate geospatial data stream processing with QoS guarantees. We have assessed the system’s performance, accuracy, and scalability by testing it in real-world application scenarios. We have designed our experiments in such a way that our evaluation is not only reporting metrics, but also validating the main architectural design decision of EdgeApproxGeo, for which we concluded that edge-based, decentralized, geohash-based stratified-like sampling performed in edge computing nodes (specifically our sampling method EdgeSOS) significantly reduces cloud-side data load time, in addition to end-to-end latency; while at the same time preserving statistically meaningful query results. We also discuss in detail the system performance across multiple aspects, including Kafka throughput under data sampling pressure, the computational efficiency of the EdgeSOS algorithm on resource-constrained edge computing nodes, and the statistical accuracy of AQP aggregates. Our aim is to provide a comprehensive understanding of the trade-offs between accuracy, resource consumption, and latency. The insights we report here are essential for interested practitioners who may seek to deploy adaptive, QoS-aware, geospatial data stream analytics pipelines in bandwidth-aware and latency-sensitive IoT environments, such as in complex smart city scenarios.

\subsection{Testing setup}

\subsubsection{Setup}

We conducted our experiments with two deployment settings to evaluate EdgeApproxGeo on varying operational configurations and contexts as follows:
\begin{enumerate}
    \item In the first deployment setting, we created an on-premises docker-based containerized deployment aiming at assessing time-based performance metrics (i.e., throughput and end-to-end latency) under controlled, reproducible deployment conditions. This setting environment encompasses four Docker containers, one hosting Spark Master and one hosting Spark Worker, in addition to two hosting Kafka broker nodes (managed by a lightweight ZooKeeper instance). Input IoT data derived from the Shenzhen electric taxi GPS dataset was replayed from local CSV files into dynamically created Kafka topics, aiming at simulating real-time smart city big geo-referenced data streaming. The continuous query running and results visualization were managed using Apache Spark's interactive Jupyter interface, which consumes the Kafka data streams and applies the EdgeApproxGeo pipeline to compute interactive geo-spatial aggregates (e.g., mean speed per neighborhood, and Top-N congested zones in the city) under defined Quality-of-Service (QoS) constraints. This lightweight, containerized deployment setup is used to measure the processing latency and throughput for the main edge-cloud workflow of our system.
    \item In contrast, the second deployment setting is a production-grade Cloud-based deployment using Microsoft Azure, aimed at testing the scalability. We have provisioned a dedicated Microsoft Azure virtual network that comprises two separate HDInsight clusters, configured so that they operate via direct, low-latency communication. The analytics cluster is hosting Apache Spark and is responsible for executing the Spark-based geospatial data processing pipeline of our system. It was provisioned with Apache Spark 3.5.2 and comprised nine nodes: two head nodes (Azure D12 v2: 4 cores, 28 GB RAM each), four worker nodes (Azure D13 v2: 8 cores, 56 GB RAM each) and three ZooKeeper coordination nodes (Azure A2 v2: 2 cores, 4 GB RAM each). A separate Apache Kafka cluster was provisioned to handle the data stream ingestion and routing. Input data was stored in CSV file format within an Azure Blob Storage container (or equivalently, a GitHub repository). To simulate real-time big geospatial data streaming conditions, data from the CSV files were replayed into Kafka topics. Continuous query running and result visualization were managed using interactive Jupyter notebooks, connected to Spark, and consuming the Kafka data streams while applying the EdgeApproxGeo processing pipeline. This native cloud deployment is used to benchmark our system performance against cloud-only baselines, in addition to evaluating scalability, while measuring the corresponding gains in terms of reductions in end-to-end latency. We also captured the accuracy-based figures (i.e., MAE and MAPE) using the same settings for each deployment and dataset independently.
\end{enumerate}

\subsubsection{Datasets}

The evaluation uses two geo-referenced datasets (vehicle mobility and hyperlocal air quality).

The first is a real-world mobility dataset that captures the trajectories of around 664 electric taxis operating in the city of Shenzhen, China. It consists of roughly 1,155,653 georeferenced tuples.
Each record contains vehicle ID, timestamp, latitude, longitude, and instantaneous vehicle speed \cite{wang2019experience}. The data set represents a real-world high-speed, big data stream of urban mobility, which is used to testing the spatial sampling and routing mechanisms of EdgeApproxGeo in the on-premises deployment.

The second is the Chicago Air Quality Dataset (collected from Project Eclipse). We used it to evaluate EdgeApproxGeo in a cloud-hosted environment. This is a real-world environmental sensing dataset captured from the Project Eclipse network~\cite{daepp2022eclipse}. Project Eclipse is a low-cost hyperlocal air quality sensing platform that was developed by the Urban Innovation Group at Microsoft Research and was designed specifically to provide significant increases (10$\times$--100$\times$) in geographic and temporal granularity for urban environmental monitoring. We used only a part of this dataset, approximately 129,532 georeferenced records collected across the city of Chicago, each tuple containing various fields, such as sensor identifier, timestamp, latitude, longitude, and PM$_{2.5}$ concentration readings. This dataset provides a real-world, spatially-skewed stream of environmental IoT data, which we used for benchmarking the scalability and network-efficiency of EdgeApproxGeo edge-cloud architecture in the cloud-hosted deployment setting that we previously described.

\subsubsection{System configuration}
Unless explicitly stated otherwise, we configured the system to use a 6-character Geohash precision for spatial discretization. This level of geohash granularity was selected because it is known to strike a balance between spatial resolution and computational efficiency, thus enabling the formation of statistically meaningful strata for the EdgeSOS algorithm of our system, while remaining feasible for real-time processing on edge computing devices.

\subsubsection{Evaluation objectives}
In testing our system, EdgeApproxGeo, we mainly focus on three performance dimensions as follows: (1) data ingestion and routing throughput. We aim to measure the end-to-end performance of the Apache Kafka-based geospatial data ingestion pipeline, including the throughput and latency of message production, consumption, and routing by varying loads and sampling fractions. (2) Efficiency of the analytics engine. We aimed at evaluating the computational performance of the Apache Spark cluster, specifically the running time and resource utilization for processing approximate geo-spatial aggregates from the pre-sampled, spatially-partitioned big data streams. (3) Approximation accuracy. We aim to quantify the statistical precision of the AQP approximate results, by measuring error metrics (e.g. MAPE, MAE) against a ground-truth baseline (i.e. 100$\%$ sampling). This test includes an analysis of measuring how accuracy is affected by main parameters such as sampling fraction and Geohash precision.

This comprehensive testing approach ensures that our evaluation captures the practical trade-offs between efficiency, latency, and result's accuracy, which are central to the value proposition of our system EdgeApproxGeo.

\subsection{Performance evaluation of edge nodes}

In this sub-section, we present a detailed, multi-aspect analysis covering the computational efficiency, throughput, and statistical accuracy of the edge-based components in our system EdgeApproxGeo. The evaluation focuses on three main performance aspects that are essential to sense the feasibility and utility of edge-based spatial AQP. Those three aspects are the following: (1) we measure the end-to-end throughput of the Kafka-based data stream ingestion, partitioning, and routing pipeline, (2) we measure the computational latency and scalability of our EdgeSOS stratified-like sampling algorithm, and (3) we measure the statistical accuracy of AQP aggregates computed at the Cloud-based modules, basically as a function of sampling rate and geospatial granularity. 

We posit that those metrics synergistically have utility in determining whether edge-based data stream preprocessing is possible to be executed in a plausible balanced way, that does not introduce extra layer of complexity, and with sufficiently low overhead, in order ultimately to justify its deployment and utility in latency-sensitive, resource-constrained environments for complex smart city scenarios.

\subsubsection{Kafka throughput and batch optimization}

\begin{figure}[h]
\centering
\includegraphics[width=0.90\linewidth]{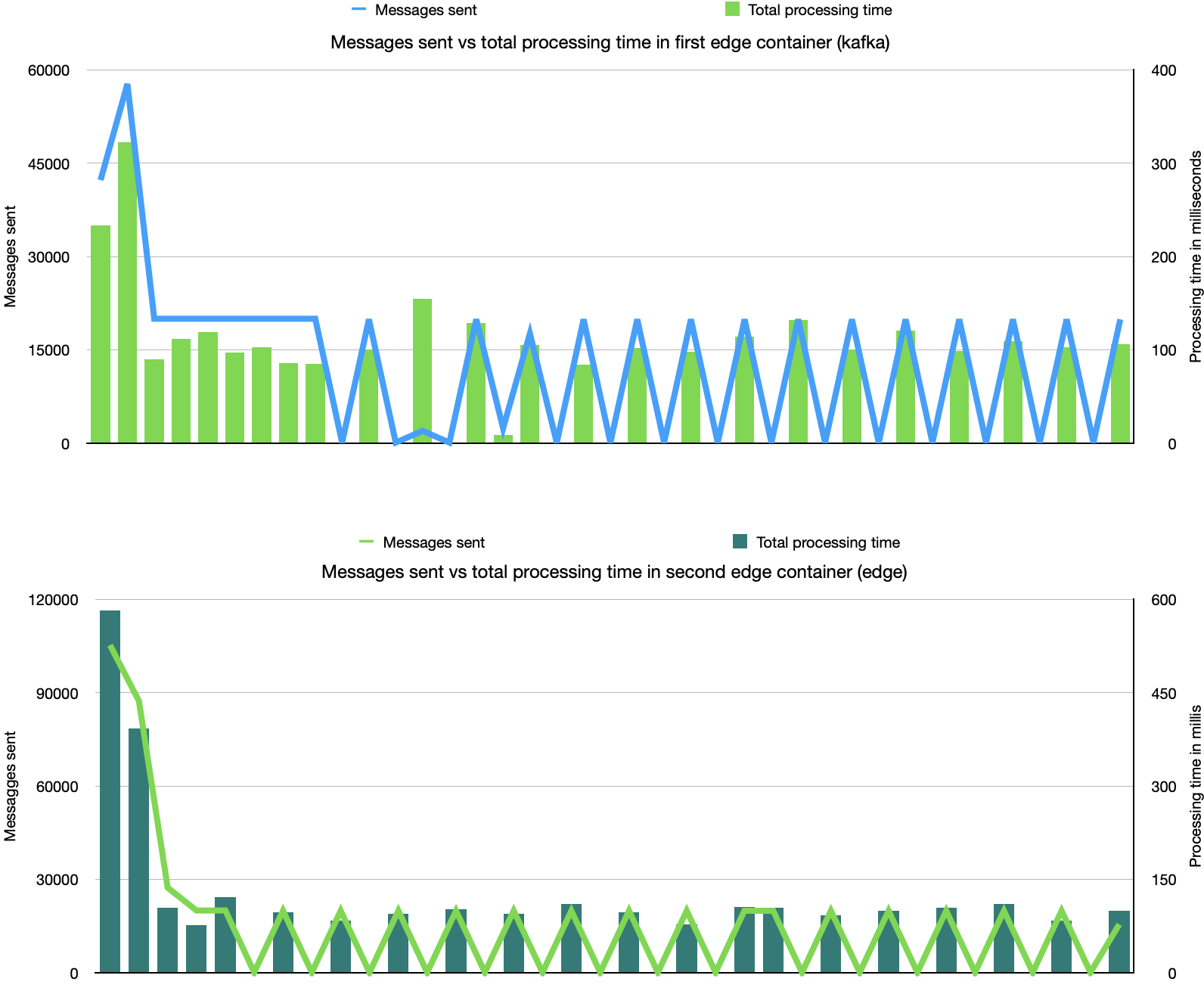}
\caption{Time-series view of messages sent (blue line) and total processing time (green bars) for the first edge container ("kafka"). After an initial transient spike, the system stabilizes at approximately 15K--20K messages per batch with around 100 ms latency, thus confirming the efficiency of the ingestion layer. Bottom chart shows a time-series view of messages sent (green line) and total processing time (teal bars) for the dedicated second edge container ("edge"). The node demonstrates stable, low-variance processing after an initial warm-up phase, highlighting the benefit of isolating sampling tasks from I/O-heavy ingestion duties.}
\label{fig:kafka_throughput}
\end{figure}

Theoretically speaking, Apache Kafka is capable of ingesting data stream rates that exceed one million messages per second; however, the practical throughput in EdgeApproxGeo is governed by the interplay between message serialization, sampling computation, and client library efficiency. Our experiments identify two primary bottlenecks that affect the overall performance of the system as a whole. These are as follows: (i) the computational cost of the sampling operation, which scales with batch size, and (ii) the performance ceiling imposed by the Kafka client library (\texttt{kafka-rust}) that is used in our Rust-based edge binaries.

to factor out the impact of sampling on Kafka throughput, we first conducted a baseline test without sampling. In other words, this baseline experiment isolates the ingestion layer to quantify the network and serialization overhead. As shown in Figure~\ref{fig:kafka_throughput}, the system maintains a stable throughput of approximately 20,000 messages with a consistent processing latency of around 100 ms. We noticed that data stream batches ranging from 15,000 to 20,000 messages incur almost identical data transmission latencies (around 100 ms), thus indicating that fixed per-batch overheads (e.g. network round-trip and serialization) mainly dominate the overall cost for smaller payloads. As a consequence, we observed that a batch size of around 20,000 messages can maximize throughput by compensating for those fixed costs over a larger number of data records, thus ultimately achieving an effective data rate approaching a theoretical maximum of $\sim$200,000 messages per second under ideal setup and operation conditions.

These results are heavily based on the specific client library and deployment environment that were used. In future work, we would evaluate the use of alternative Kafka clients (e.g. rust-rdkafka) to conclude if it is possible to achieve higher baseline throughput, thus relaxing constraints on the budget assigned to sampling computation. We have also noticed that adaptively tuning the sampling window size (i.e., the batch interval at which buffered messages are processed and forwarded downstream) has a utility in managing load. By configuring the time interval window to accumulate approximately 20,000 messages before triggering sampling and transmission, our system achieves high throughput while keeping per-batch latency under control. This finding confirms the importance of the co-design choice where we incorporate the sampling logic with data ingestion parameters to avoid extra layers of complexity that may create a bottleneck at the edge-side.

We have also independently analyzed the dedicated sampling node to further validate the stability of our edge processing layer. As shown in Figure~\ref{fig:kafka_throughput}, the performance of the dedicated edge container is predictable despite a larger initial transient cost (i.e., approximately 115K messages / 580 ms), it quickly stabilizes to process around 15K--25K messages per batch with low variance in processing time (approximately 50--150 ms). Having said that, the EdgeSOS algorithm itself is deterministic and scalable. The volatility that we observed in co-located nodes is thus attributed to resource contention.

\subsubsection{EdgeSOS sampling latency and scalability}\label{EdgeSOSSampling}

The usability of our system EdgeApproxGeo heavily relies on the ability of our EdgeSOS edge-based sampling method to execute geohash-based stratified sampling within the latency budget determined by the Kafka batch interval. A single-thread implementation is considered inadequate, and shows sub-optimal scaling that makes it unusable for data stream batches that exceed 50k tuples. To mitigate this, we chose to implement a parallel version of our sampling method using the Rust language rayon crate to distribute the workload of sampling individual strata to threads in a thread pool.

\begin{figure}[h]
\centering
\includegraphics[width=0.90\linewidth]{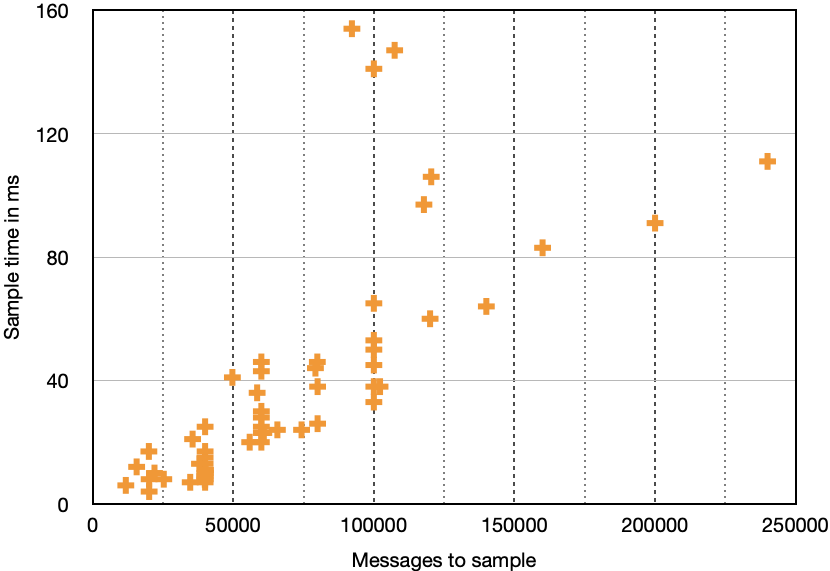}
\caption{EdgeSOS sampling latency as a function of data stream input size. Parallel execution (using Rust rayon) ensures near-linear scaling, thus enabling real-time processing of batches up to 100K tuples.}
\label{fig:sampling_latency}
\end{figure}

As shown in Figure~\ref{fig:sampling_latency}, this parallel design enables our system to achieve a near-linear monotonic scaling by varying the input size. In more detail, for a batch size of 100k data messages, the sampling latency is reduced by a factor of 4 compared to the baseline, confining it closely within the 100 ms target time interval, which is a utility established by our Kafka throughput and batch optimization. The aggregate results across both edge nodes, as shown in Figure~\ref{fig:sampling_latency}, which depicts consistency, with latency increasing monotonically as a quasi-linear function of the tuple count of the data. We observed a small number of outliers (three samples around the 100k tuple mark), which is probably caused by the fact that the edge node concurrently handles data distribution tasks, causing a possible resource contention.

Another important observation is that sampling latency was found to be largely independent of the sampling fraction (e.g. 20\% vs. 80\%). This is attributed to the fact that EdgeSOS’s cost is mainly dominated by the initial grouping of data stream messages into geohash strata, and the subsequent SRS sampling within each stratum individually has a computational cost dominated by the total number of arriving messages, and not the fraction of data kept after sampling. This property allows us to adaptively adjust sampling fractions in our system for the sole purpose of QoS and scarce resource management, without introducing unexpected latency spikes.

\begin{figure}[h]
\centering
\includegraphics[width=0.90\linewidth]{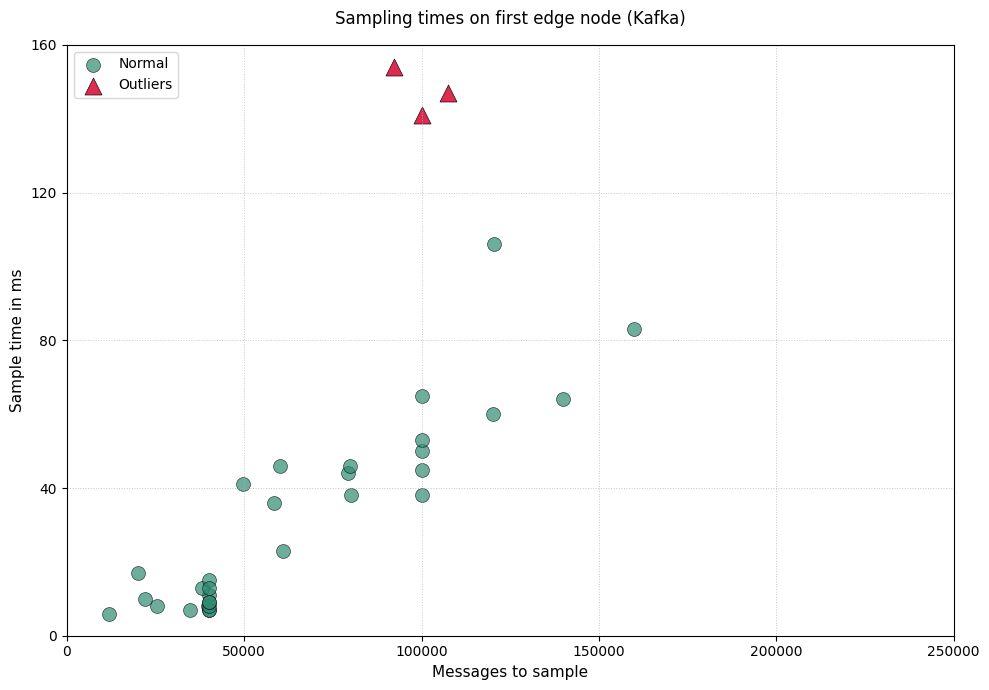}
\caption{Sampling latency on the edge node co-located with the data distribution service (named "kafka"). The three high-latency outliers (at $\sim$100K messages) are attributed to resource contention from concurrent duties.}
\label{fig:sampling_latency_kafka_node}
\end{figure}

\begin{figure}[h]
\centering
\includegraphics[width=0.90\linewidth]{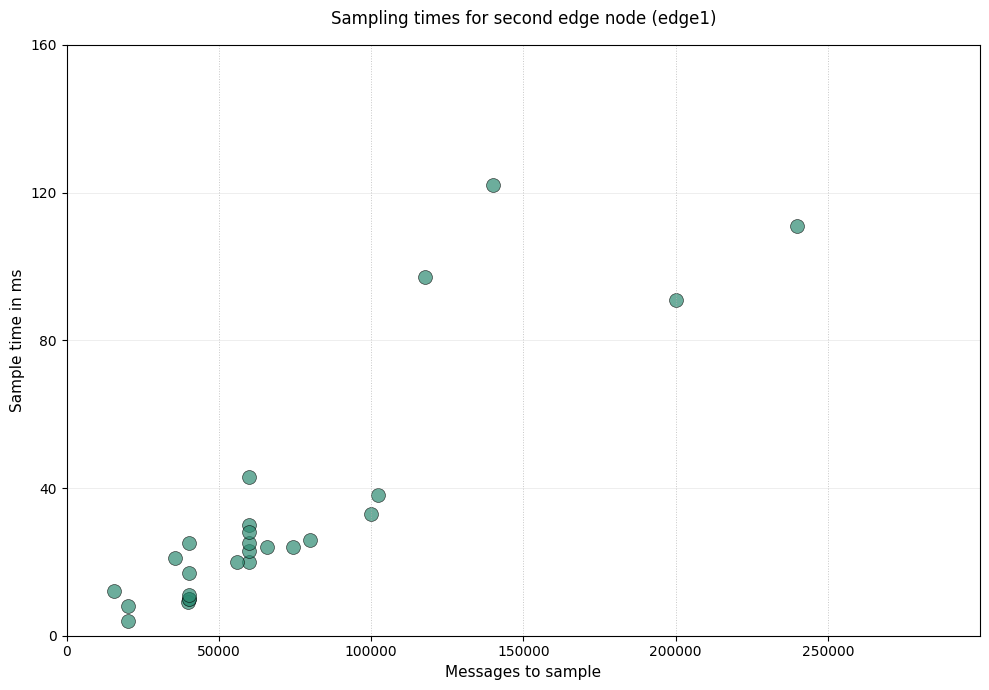}
\caption{Sampling latency on the dedicated edge processing node (named "edge1"). Performance is stable, which confirms that the sampling algorithm itself is not the source of variability.}
\label{fig:sampling_latency_edge_node}
\end{figure}

To further investigate the source of the latency outliers that we observed earlier, we performed an analysis on the performance of the sampling algorithm on each edge computing node individually. Figure~\ref{fig:sampling_latency_kafka_node} demonstrates the sampling latency for the node co-located with the data distribution service (labeled as "kafka"). This edge node has the responsibility of IO-intensive duty of consuming data from Kafka topics and the compute-intensive duty of data sampling. The three samples that cause the high-latency around 100K messages input size are clearly visible in this figure, co-located with time intervals that exhibit peaks in the data stream load. In contrast, Figure~\ref{fig:sampling_latency_edge_node} demonstrates a latency profile for a dedicated edge processing node (which is labeled as "edge1"), a node that exclusively performs data sampling. The difference in this case is clear as the latency curve is smooth,  monotonic and exhibits better near-linear scaling that does not show significant outliers. This comparison demonstrates that the EdgeSOS algorithm itself is scalable and predictable. The observed performance variance (i.e., outliers) in the aggregate results (i.e., both edge nodes) is a direct consequence of deployment topology and resource competition, and the core sampling logic is thus efficient. This indicates that performance can be optimized in such deployments by isolating resources (i.e., dedicating nodes to perform specific tasks) rather than requiring fundamental algorithmic changes.

\subsubsection{Accuracy and visual accuracy of approximate results}
\begin{figure}[h]
\centering
\includegraphics[width=0.90\linewidth]{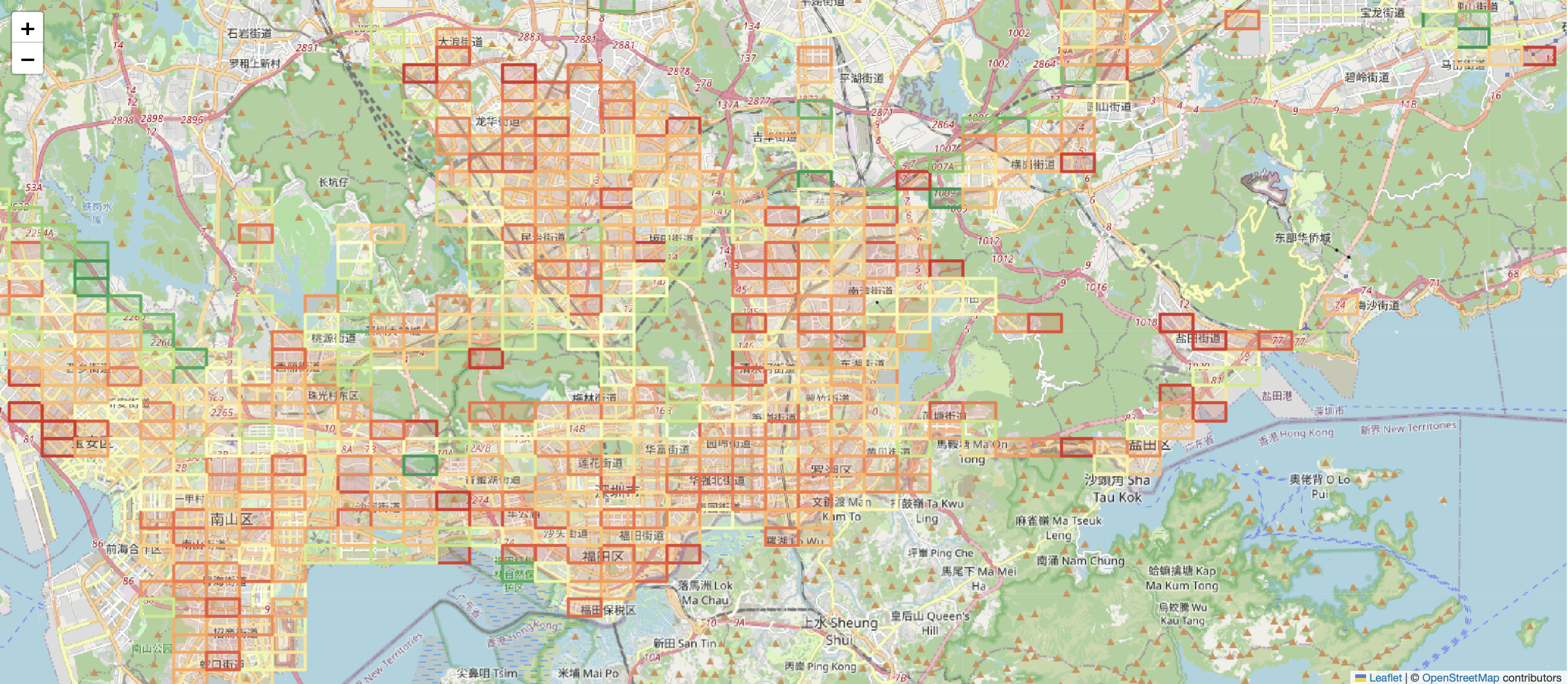}
\caption{Baseline heatmap at 100\% sampling (ground truth).}
\label{fig:heatmap_100}
\end{figure}

\begin{figure}[h]
\centering
\includegraphics[width=0.90\linewidth]{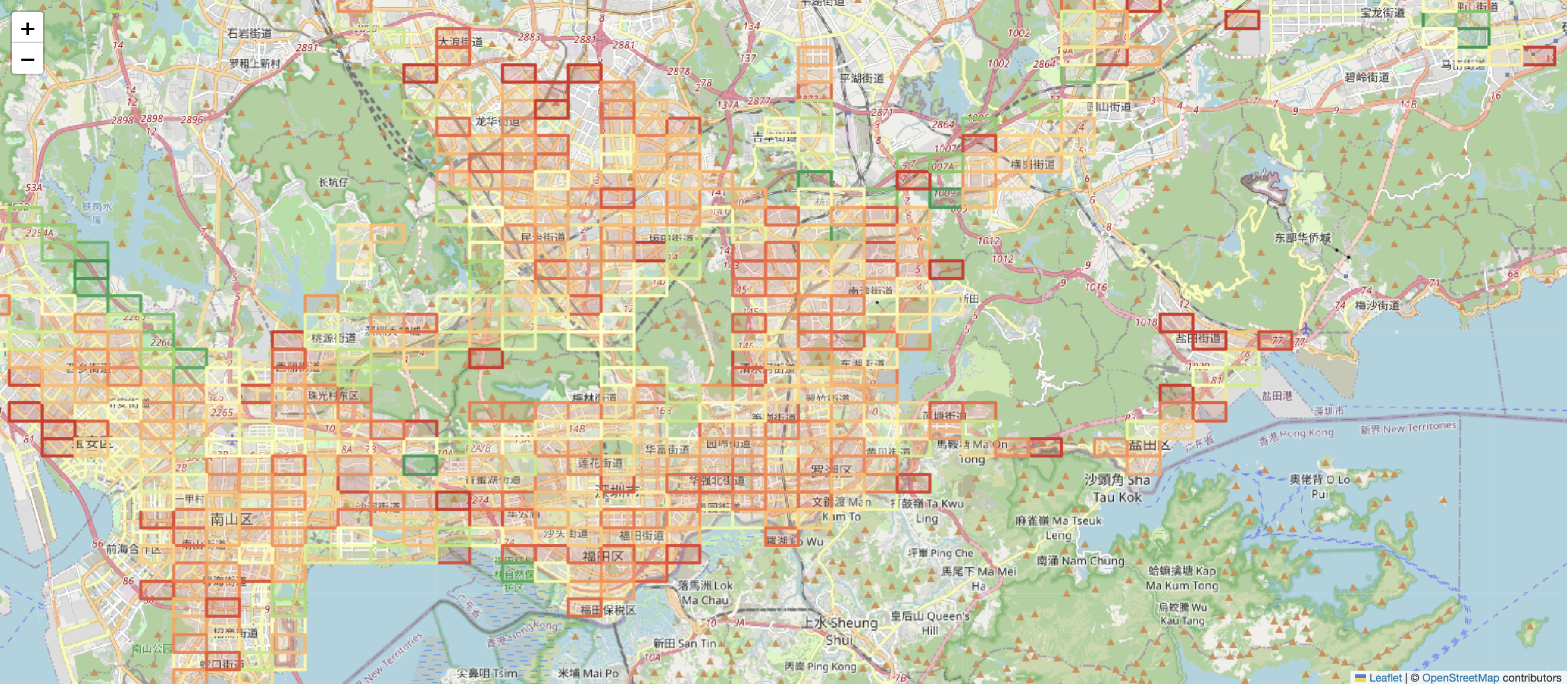}
\caption{Heatmap at 80\% sampling. Visually and statistically (MAPE$<$10\% ) nearly identical to the baseline, suitable for operational use.}
\label{fig:heatmap_80}
\end{figure}

\begin{figure}[h]
\centering
\includegraphics[width=0.90\linewidth]{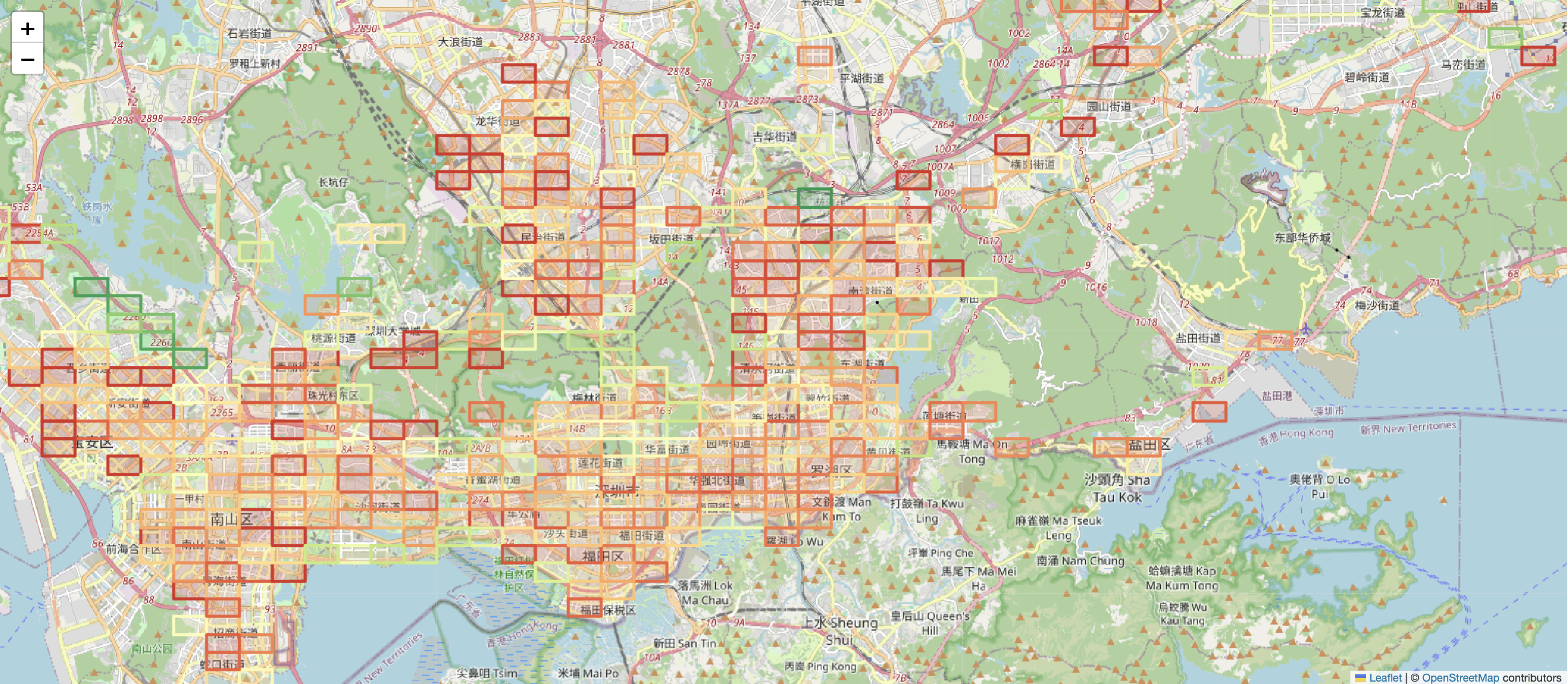}
\caption{Heatmap at 20\% sampling. While individual geohash estimates show higher variance (MAPE$\approx$38\% ), macroscopic traffic patterns (e.g., major congestion zones) are preserved, enabling trend analysis.}
\label{fig:heatmap_20}
\end{figure}

\begin{figure}[h]
\centering
\includegraphics[width=0.90\linewidth]{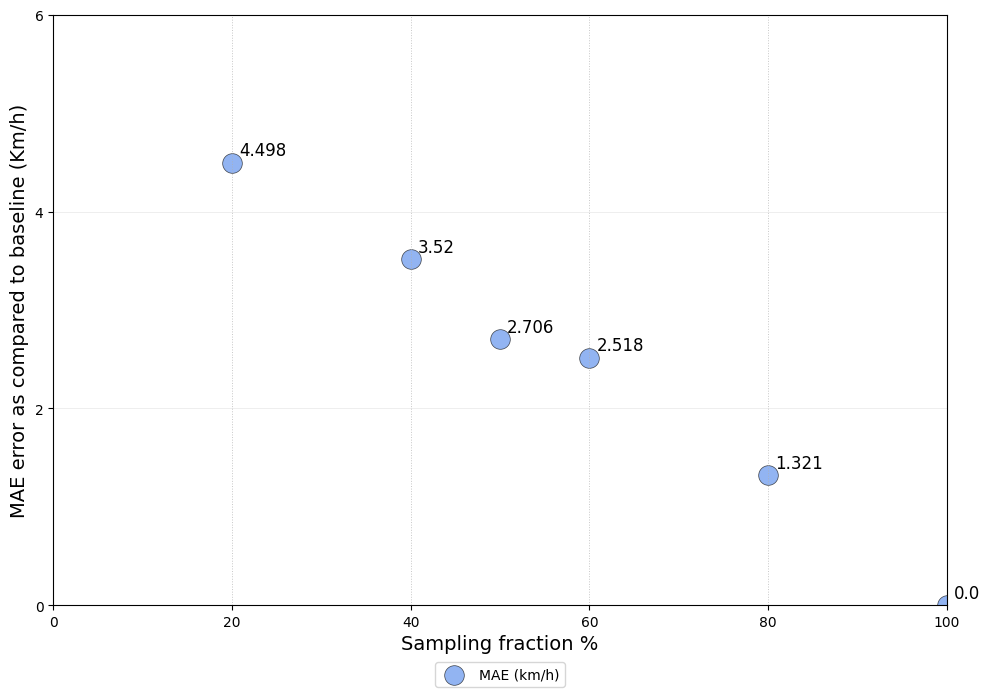}
\caption{Mean Absolute Error (MAE) decreases near-linearly as we increase the sampling fractions, in accordance with sampling theory, as expected.}
\label{fig:mae}
\end{figure}

\begin{figure}[h]
\centering
\includegraphics[width=0.90\linewidth]{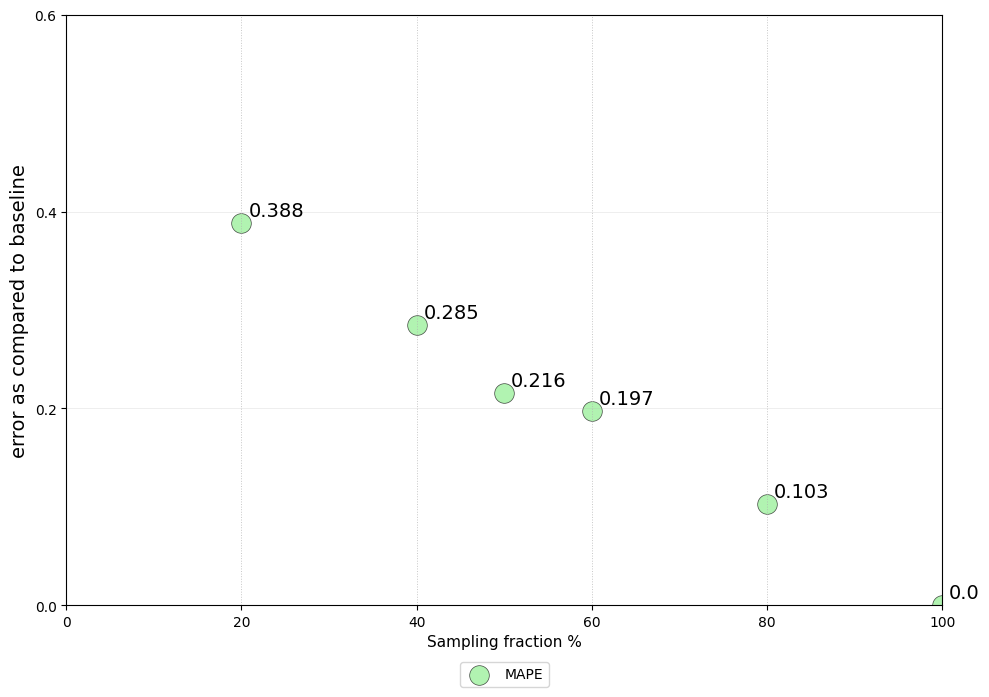}
\caption{Mean Absolute Percentage Error (MAPE) as a function of sampling fraction. Error remains below 10\% for sampling fractions of 80\% and above.}
\label{fig:mape}
\end{figure}

\begin{figure}[h]
\centering
\includegraphics[width=0.90\linewidth]{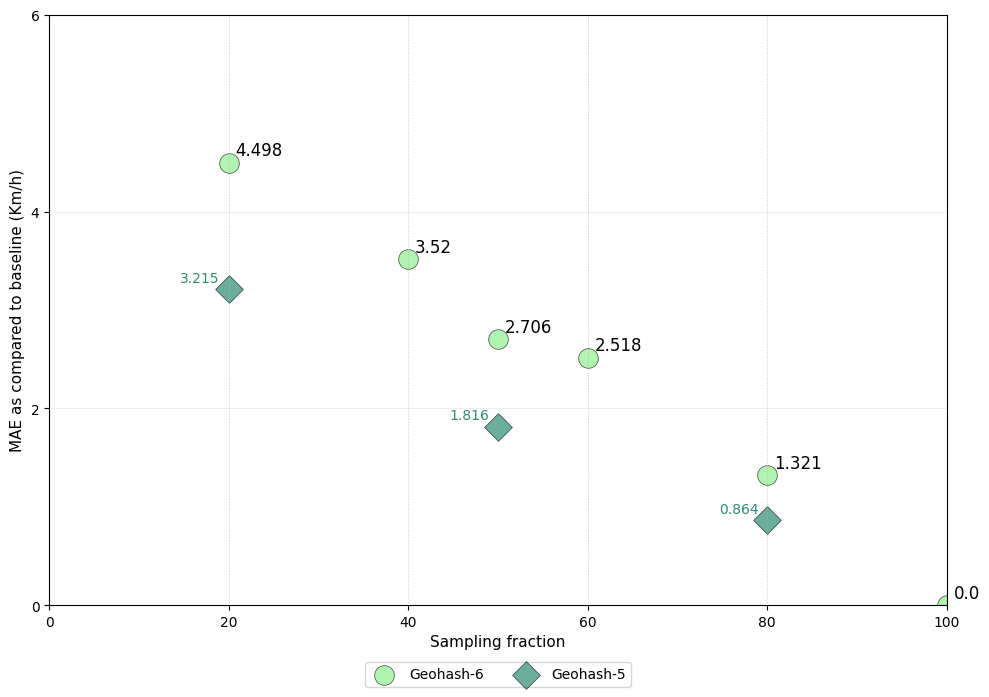}
\caption{Comparison of MAE between Geohash-5 and Geohash-6. Geohash-5 consistently results in lower absolute errors mainly due to larger, more statistically stable strata.}
\label{fig:mae_geohash5_vs_6}
\end{figure}

\begin{figure}[h]
\centering
\includegraphics[width=0.90\linewidth]{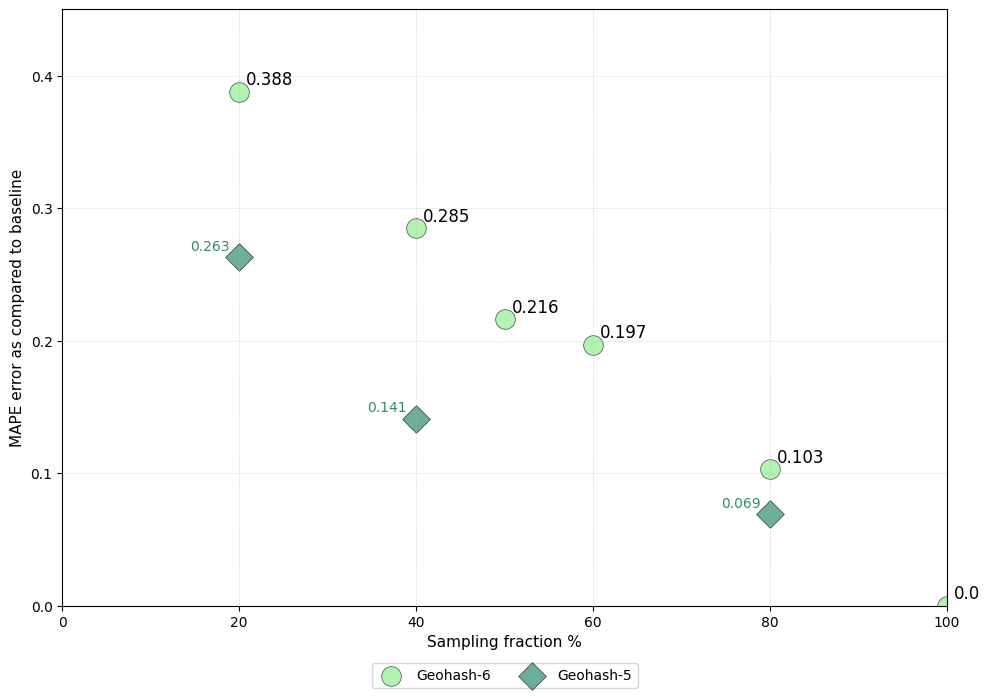}
\caption{Comparison of MAPE between Geohash-5 and Geohash-6. At 80\% sampling, Geohash-5 (7\%) outperforms Geohash-6 (10\%), demonstrating the accuracy benefit of coarser spatial granularity.}
\label{fig:mape_geohash5_vs_6}
\end{figure}

We evaluated the statistical accuracy of EdgeApproxGeo AQP aggregates. Using the Shenzhen taxi dataset that we described previously, we calculated the Mean Absolute Percentage Error (MAPE) and Mean Absolute Error (MAE) of the average vehicle speed queries and captured values by varying sampling fractions (20\%–100\%) and using Geohash 6 as strata. The key findings are summarized below and visualized in Figures~\ref{fig:heatmap_100}-\ref{fig:mape_geohash5_vs_6}:
\begin{itemize}
    \item High precision at 80\% Sampling fraction. At an 80\% sampling fraction, the MAPE value remains approximately 10\% (as shown in Figure~\ref{fig:mape}), which is a level of error that is typically acceptable for operational decision-making. The heatmap generated from the 80\% sample (as shown in Figure~\ref{fig:heatmap_80}) is virtually indistinguishable from the ground-truth baseline counterpart (as depicted in Figure~\ref{fig:heatmap_100}), a significant finding that further confirms that global spatial patterns are preserved
    \item Accuracy at low sampling fractions (specifically 20\%). Even at an aggressive 20\% sampling fraction, while MAPE increases to about 38\% (as shown in Figure~\ref{fig:mape}), the system maintains its property in handling high-level trend analysis (e.g., via heat maps). For example, as shown in Figure~\ref{fig:heatmap_20}, city-wide congestion hotspots and traffic flow patterns remain clearly distinguishable, demonstrating that EdgeApproxGeo can provide actionable insights even with stringent resource constraints
    \item We observed a correlation between geohash granularity and results accuracy, with a clear trade-off visible in our results. We observed how coarser spatial granularity improves the statistical accuracy of query results. For example, when using Geohash precision 5 (with a larger cell size) instead of Geohash precision 6, MAPE at 80\% sampling fraction drops from 10\% to 7\%, representing a relative improvement of circa 30\% compared to the finer granularity (Figure~\ref{fig:mape_geohash5_vs_6}). This basically is due to the fact that larger geohash cells typically contain more data points which leads to more accurate sample-based data statistics (i.e., means and lower variance estimations). This parameter allows the trading-off fine-grained spatial resolution for an improved statistics calculation accuracy and assists developers in calibration based on their specific QoS requirements.
\end{itemize}

The experiments show that the EdgeApproxGeo design successfully balances QoS goals (efficiency and accuracy). The system does not solely reduce data stream volume at the edge, it also preserves the geo-statistical representativeness of the original data stream and provides users with tunable and predictable error bounds.

\subsubsection{Design implications and putting it all together}

Our experimental results provide several insights for the successful design and deployment of edge-based geospatial AQP data stream systems as follows: (1) the batch size of the data stream has a direct impact on overall performance for batch-driven DSP systems (such as Apache Spark). Indeed, both Kafka throughput and EdgeSOS latency are predominantly determined by the volume of data processed for each batch. Our experimental results show that batch size of circa 20k messages optimizes the latency-throughput trade-off, which provides actionable insights for DSP system's configuration. It is therefore advisable to employ window-based processing in DSP systems if the smart city scenario requires a predictable and tunable performance. (2) Adaptive data stream time windows. Our current standard-compliant prototype features static, time-based windows for data stream processing. However, our experimental results suggest that adaptive and proactive data-driven time windows (e.g. trigger processing after a fixed number of data tuples arrive) are promising for production-grade smart city scenarios, which typically experience oscillating data traffic loads. In general, this way, the computational load per batch remains constant, thus preventing latency spikes during unpredictable surges in data traffic. (3) Granularity as a tunable parameter. Our results show that Geohash-5 outperforms Geohash-6 in accuracy; thus, spatial granularity is not solely a resolution setting, but also an essential accuracy-efficiency balancer. Geohash precision is therefore an important tunable parameter that designers should select based on some characteristics of the data stream being processed, the geographical area covered, and the required level of spatial detail. (4) The utility of stratified sampling in the AQP of data streams. The consistent, near-linear scaling of EdgeSOS performance, as shown in our results, in addition to its ability to keep error rates in check, even at high data reduction fractions, validates the selection of stratified sampling for AQP of geospatial data streams over conventional counterparts such as SRS. This is attributable to the fact that stratified-based schemes select proportional data representation from all spatial regions, a design intrinsic in EdgeSOS, thereby avoiding the possible flaws of under-sampling sparse regions, which is typical in non-stratified sampling approaches.

In conclusion, our experimental results on the performance of a standard-compliant prototype built upon our system EdgeApproxGeo show that our system is unique in providing QoS aware results on processing fast arriving highly skewed big geo-referenced data streams. The two main components of our system (the EdgeSOS algorithm and spatially-aware routing) are practical in complex time-sensitive resource-constrained smart city scenarios. They operate in synergy to provide a significant system operation speedup (i.e., throughput and latency QoS constraints), while providing statistically plausible, and tunable accuracy, which makes EdgeApproxGeo a high-performing and efficient architecture for real-time, resource-constrained geospatial big data analytics.

\begin{figure}[h]
    \centering
    \includegraphics[width=0.90\linewidth]{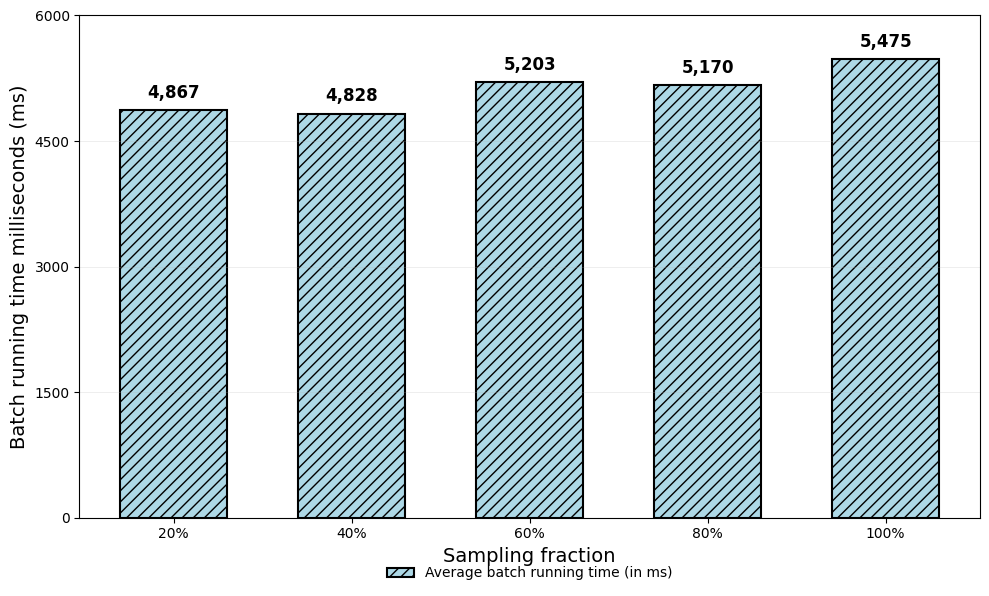}
    \caption{average batch running times change by varying sampling fractions in our tests.}
    \label{fig:spark_batch_time}
\end{figure}

\subsection{Spark performance test}

We use the Spark Web UI to measure Apache Spark performance. Several useful performance metrics are captured by the Spark UI for various tasks performed by Spark.
We measured the average duration of a Spark batch with varying sampling fractions. Specifically, we carried out several tests with sampling fractions that are set to values ranging from 20\% to 100\%, with a step of 20\%, where we always use our stratified sampling algorithm and the neighborhood-based data routing module that we discussed previously (i.e., data tuples are
distributed to dedicated topics for each neighborhood).
The results of our tests, shown in Figure \ref{fig:spark_batch_time}, demonstrate a subtle positive correlation between the sampling fraction and the average batch run time, the difference in performance captured between analyzing a data sample of 20\% versus 100\% size is subtle, where the latency decreases by approximately 11-12\% compared to a data sample of 100\%. The limited performance gain is
primarily due to the fact that the dataset size that we used for benchmarking (the Shenzhen mobility dataset in this case) did not fully stress the Spark cluster, and consequently fixed overheads (e.g., job scheduling) remain the dominant performance impact factors regardless of the data stream volume.

\subsection{Benchmarking against existing Cloud-Only solutions}
We have benchmarked our system EdgeApproxGeo against SpatialSSJP
\cite{al2023spatialssjp}, a solution employing cloud-only geohashing, neighborhood categorization, and sampling, to assess the magnitude of the performance gains derived by the ability to pre-compute those steps on edge nodes.

The experiments have been carried out on a dataset of air quality data collected in the city of Chicago in 2021. Our results show that by avoiding spatial joins in the cloud layer, EdgeApproxGeo achieves around 15\% to 20\% reduction in execution time compared to the cloud-only SpatialSSJP baseline (benchmarking performed using the Chicago AQ dataset and the experiment was deployed on Microsoft Azure), depending on the run and the sampling percentage, while retaining very similar statistical properties compared to carrying out stratified sampling in the cloud nodes.

This means that, by pre-processing data on edge-cloud nodes, we can not only reduce the amount of data that is transferred, thus reducing networking costs, but also significantly speed up the post-processing of data on the cloud, and thus cutting the execution time and the amount of resources needed for the cloud nodes, which can further reduce costs. Moreover, doing all of this is possible without significantly sacrificing the accuracy of calculations, when compared with SpatialSSJP.

In the following discussion, we refer to computing on the cloud with the full dataset as the \textit{baseline}, while cloud-sampled will refer to SpatialSSJP, and lastly edge-sampled will refer to EdgeApproxGeo.

\begin{figure}[h]
\centering
\includegraphics[width=1\linewidth]{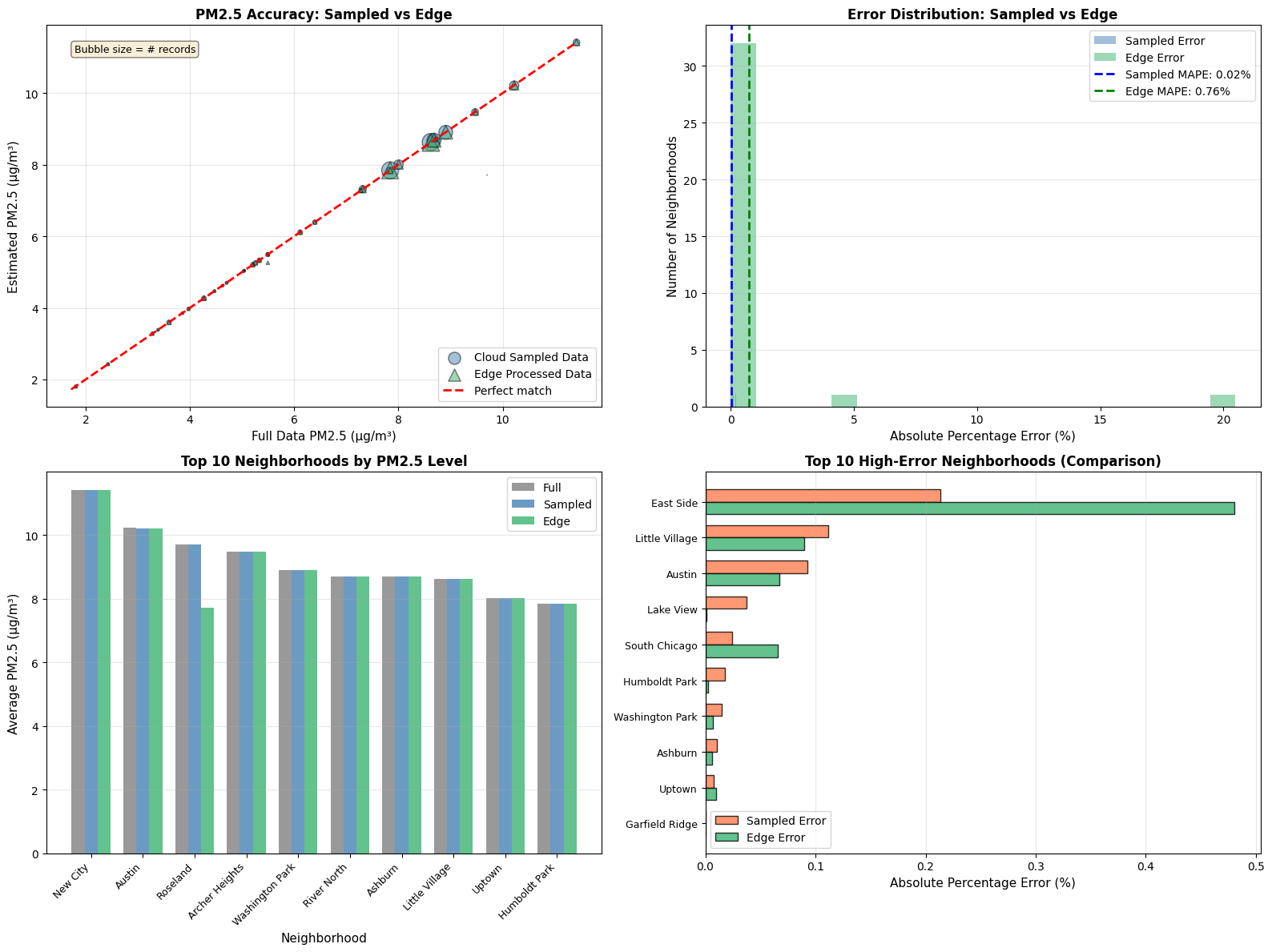}
\caption{Comparison of statistical errors: Baseline (full data), Cloud-Sampled (SpatialSSJP), and Edge-Sampled (EdgeApproxGeo). The data refers to the baseline (or full) to performing tests on the full dataset, while SpatialSSJP with sampling at 80\% carried out on the cloud is called sampled, and our system EdgeApproxGeo with sampling at 80\% is referred to as edge.}
\label{fig:baseline_vs_edge}
\end{figure}

As shown in Figure~\ref{fig:baseline_vs_edge}, which compares the performance of SpatialSSJP and EdgeApproxGeo with a sampling fraction of 80\% of the total amount of data, we can see that for SpatialSSJP the average neighborhood error is approximately 0.02\%, with peaks around 0.22\%, while for most neighborhoods the error remains within 0.10\%. In contrast, EdgeApproxGeo shows a slightly higher error rate of 0.76\%, with almost all neighborhoods staying below that threshold, and a few outliers, including errors of 5\% and 20\%.

to facilitate interpretation of results shown in Figure~\ref{fig:baseline_vs_edge}, the top-left scatter plot shows that both methods (i.e., Edge-based and Cloud-based) cluster tightly around the ground truth (represented as the red dashed line in the figure). The top-right histogram shows the global error distribution and depicts the low MAPE achieved by both systems despite the lack of edge outliers. The bottom-left bar chart demonstrates that for high-density (a.k.a. hotspot) neighborhoods, the Edge and Cloud estimates are almost visually identical to the full data baseline. Finally, the bottom-right chart isolates the 10 neighborhoods with the highest error, demonstrating that even in the sparse regions of the worst-case, the absolute error remains manageable (less than 0.5\% for typical cases, with rare extreme outliers excluded from this specific view for clarity).

The slightly higher error rate and the presence of outliers can be explained by the edge-based solution performing periodic sampling on subsets of the overall data set, while the cloud-based solution performs the sampling in one pass on the full data set. For neighborhoods with few data points, sampling once after receiving the full dataset produces statistically more accurate averages than periodic sampling.

However, given sufficient data volume, we can see that both sampling methods yield near-identical results. To limit the likelihood of outliers, one can think of adding optimizations to the edge sampling strategy, such as avoiding sampling neighborhoods with too few data points in a given sampling window, as the benefit of reduced bandwidth is outweighed by the potential loss in precision for those specific neighborhoods.

In conclusion, the MAPE and the absolute percentage error (APE) (which measure how far the estimate was from the values obtained in the ground truth) for both SpatialSSJP and EdgeApproxGeo, excluding the outliers, show no significant differences, which means that carrying out the sampling on edge nodes as opposed to doing it in the cloud does not lead to a significant difference in accuracy.

Furthermore, to validate the time efficiency gains of our proposed edge-cloud architecture, we compare it with SpatialSSJP~\cite{al2023spatialssjp}, a state-of-the-art cloud-only geospatial big data AQP stream processing system. This comparison aimed to quantify the time-based performance gains achieved by distributing sampling and computation between edge computing nodes and cloud-based resources compared to centralized cloud-only processing counterparts.

\begin{figure}[t]
    \centering
    \includegraphics[width=\linewidth]{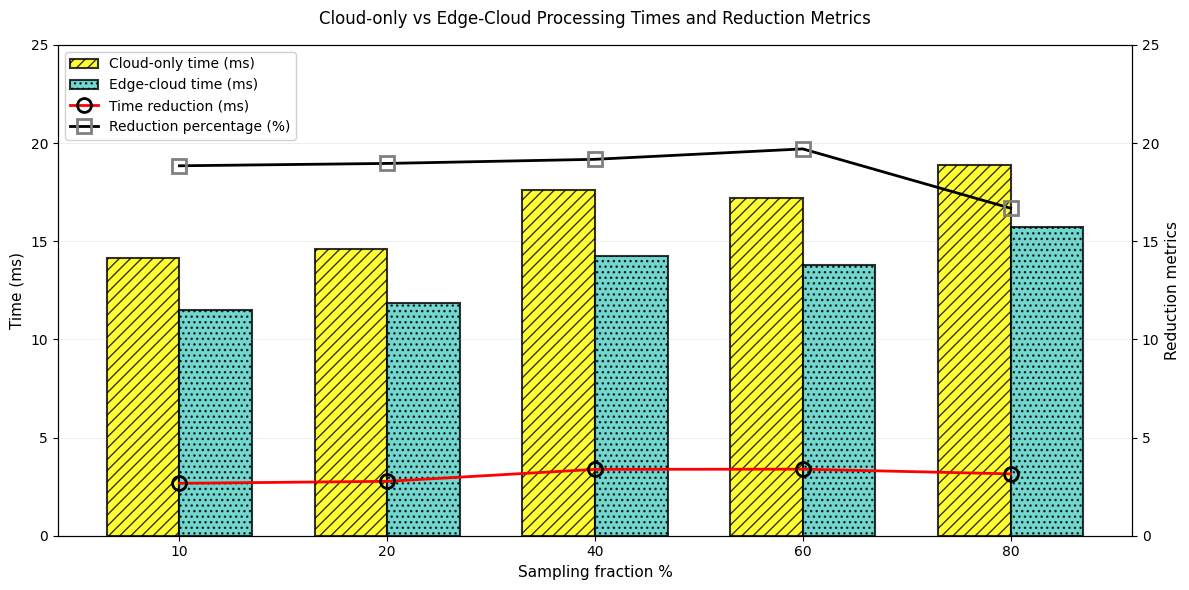}
    \caption{Comparison of processing times between cloud-only (SpatialSSJP) and edge-cloud approaches across different sampling fractions. The edge-cloud architecture consistently achieves lower processing times, with reduction percentages ranging from approximately 15\% to approximately 20\%.}
    \label{fig:cloud_vs_edge_cloud}
\end{figure}

Figure~\ref{fig:cloud_vs_edge_cloud} shows the time-based results that we obtained as it shows the processing time comparison between SpatialSSJP in the cloud alone as opposed to our edge-cloud architecture (EdgeApproxGeo) by varying sampling fractions (from 10\% to 80\%). The results show that our edge-cloud system outperforms the cloud-only counterpart for all sampling configurations. It is also worth noting that our edge-cloud architecture achieves a reduction in processing time that ranges from approximately 15\% to approximately 20\% compared to the cloud-only counterpart. As the sampling fraction increases from 10\% to 80\%, both approaches (Cloud-only versus our edge-cloud system) show increased processing times, a behavior that is expected and attributed to bigger data stream volumes. Nevertheless, our edge-cloud system maintains its performance advantage, with time reduction gains becoming more pronounced at higher sampling fractions.

The time reduction percentages shown in Figure~\ref{fig:cloud_vs_edge_cloud}, demonstrate that our edge-cloud system achieves a maximum time performance gain (approximately 20\% reduction) at sampling fractions that fall between 40\% and 60\%. This suggests that our distributed processing mechanism is very effective in scenarios of moderate to high sampling fractions, which are common in real-world IoT applications that require timely big data analytics.

The results presented in this paper support our conclusion that big data analytics performance improves significantly by distributing spatial big data stream processing tasks between edge and cloud computing resources, as opposed to centralized, cloud-only approaches such as SpatialSSJP. The edge-cloud architecture that we presented in this paper thus  reduces processing, latency and also provides a more scalable and network-efficient solution for real-time big spatial data streams analytics in IoT environments of advanced smart city applications.

\clearpage

\section{Conclusion and Future Research}\label{sec6}
In this paper, we introduce our system EdgeApproxGeo, as a new edge-cloud architecture designed specifically to support approximate geospatial analytics over fast-arriving geo-referenced data streams. Our system features an edge-based core stratified-like sampling component (i.e. EdgeSOS) that aims at decentralizing stratified sampling and implementing spatial-aware data stream routing at edge nodes. By this design, our approach aims at alleviating cloud-side computational load, and thus reducing the end-to-end latency while keeping statistical error margins in check.
Our extensive experimental evaluation, which we conducted using real-world big geospatial-referenced datasets (including Chicago USA low cost hyperlocal air quality and electric taxi trajectories from Shenzhen in China), and has been tested on a standard-compliant prototype that is implemented atop de facto standards of big data stream ingestion and Cloud-based big data analytics systems, specifically Apache Kafka and Apache Spark, consequently. The results show that EdgeApproxGeo achieves significant processing speedups of approximately 15\% -- 20\% compared to cloud-only baselines such as the state-of-the-art SpatialSSJP~\cite{al2023spatialssjp} . This was possible by offloading some costly in-cloud spatial processing to edge devices near the IoT data sources. Another important observation is that, in terms of the end-to-end latency, which includes network data transfer, significant reductions have been observed under specific network conditions and sampling fractions. Also, for accuracy-based QoS goals, metrics such as MAPE show less than 10\% at 80\% sampling fraction, which is plausibly comparable to the cloud-only stratified sampling counterpart, thus validating the statistical utility of the edge-cloud system.
The results we obtained from the experimental evaluation of EdgeApproxGeo are promising, however, several limitations are planned for future work. First, the experimental evaluations were conducted on controlled Cloud cluster configurations (i.e., Azure HDInsight or on-premises containerized cluster) and AQ and mobility big geo-referenced datasets. However, performance characteristics may vary in heterogeneous edge environments with harsher resource constraints or different data distributions. Second, the current implementation is based on the semantics of the tumbling window data stream processing and the stream-static join operations. However,  sliding windows data stream processing semantics and stream-stream geospatial join processing typically introduce additional state management complexities and requirements that were not addressed in this version of our system. Third, the accuracy of the results obtained in the system is heavily dependent on the chosen geohash precision, where coarser granularities may improve stability but reduce spatial resolution, thus requiring careful parameter tuning for domain-specific application contexts in advanced urban analytics and smart city scenarios. Finally, the adaptive loop feedback mechanism for tuning the sampling fractions is manually configured by expert input in our experiments; however, the architectural design of our system supports automation of such an important aspect.
Building on the findings of this article, our future research will focus on several relevant aspects, including the following: (1) We focus on designing a dynamic sampling configuration adaptation strategy, by fully automating the loop feedback mechanism (Section~\ref{subsec4}) so that it self-tunes the sampling fractions for each edge node based on real-time resource pressure and the QoS goals declared in the continuous query Service Level Objectives (SLOs). (2) We plan to evaluate alternative geospatial data tessellation schemes (e.g., Uber's H3 and Google's S2) or hybrid approaches (for instance, Geohash and Voronoi layers), thus aiming at handling irregular urban topologies and mobility patterns of cities. (3) We aim to introduce support for sliding window semantics and stream-stream join operations to extend applicability of our system to more advanced real-time event processing scenarios. (4) We aim at testing the serverless edge integration by investigating deployment on serverless edge platforms (e.g., AWS Lambda@Edge) aiming at enabling elastic, fine-grained data analytics scaling without provisioning infrastructure manually.
Our system EdgeApproxGeo shows that edge preprocessing is an effective approach for reducing latency and resource consumption in real-time big geospatial data analytics, thus offering a practical approach for scalable IoT big multi-modal data processing in advanced and demanding smart city environments.

\section*{Declarations}

\begin{itemize}
\item Author's Contributions. Conceptualization: Isam Mashhour Al Jawarneh, Luca Foschini and Paolo Bellavista; methodology: Isam Mashhour Al Jawarneh and Lorenzo Felletti; software: Lorenzo Felletti and Isam Mashhour Al Jawarneh; validation:  Lorenzo Felletti and Isam Mashhour Al Jawarneh ; formal analysis: Lorenzo Felletti , Paolo Bellavista and Isam Mashhour Al Jawarneh; investigation: Lorenzo Felletti, Isam Mashhour Al Jawarneh, and Luca Foschini; resources: Paolo Bellavista, Isam Mashhour Al Jawarneh, and Luca Foschini ; data curation: Lorenzo Felletti and Isam Mashhour Al Jawarneh; writing—original draft preparation: Isam Mashhour Al Jawarneh, Lorenzo Felletti, Luca Foschini, and Paolo Bellavista ; writing—review and editing: Isam Mashhour Al Jawarneh, Lorenzo Felletti, Luca Foschini, and Paolo Bellavista; visualization: Lorenzo Felletti, and Isam Mashhour Al Jawarneh; supervision: Isam Mashhour Al Jawarneh, Luca Foschini and Paolo Bellavista; project administration: Luca Foschini and Paolo Bellavista; funding acquisition: Paolo Bellavista; All authors read and approved the final manuscript.

\item Acknowledgments: This work  has been partially supported by the European Union under the NRRP partnership on "Telecommunications of the Future" (PE00000001 - program "RESTART") and by the National PRIN JOULE project.

\item Data Availability Statement: The two datasets used in this paper are publicly available. The electric taxi Shenzhen datasets are accessible (as of this writing) through a website at https://guangwang.me/files/ETData.rar (last accessed April, 2026). It contains anonymized GPS trajectories from 664 electric taxis and was originally published in \cite{wang2019experience}. Use is permitted for academic research only; users of these data must cite \cite{wang2019experience} and respect the data provider's terms. The second data set represents the Chicago Air Quality Dataset from Project Eclipse~\cite{daepp2022eclipse}, and it is publicly available and can be accessed (as of this writing) through the website of Microsoft's planetary computer at https://planetarycomputer.microsoft.com/dataset/eclipse

\item Competing interests: The authors have no competing interests to declare that are relevant to the content of this article.

\end{itemize}

\noindent


\bibliography{sn-bibliography}% common bib file

\end{document}